\documentclass[12pt]{article}
\newcommand{\bq}{\begin{quotation}\noindent}
\newcommand{\eq}{\end{quotation}\medskip}
\newcommand{\be}{\begin{equation}}
\newcommand{\ee}{\end{equation}}
\newcommand{\bea}{\begin{eqnarray}}
\newcommand{\eea}{\end{eqnarray}}
\newcommand{\bc}{\begin{center}}
\newcommand{\ec}{\end{center}}
\def\tr{{\rm tr}}

\usepackage{epsf}
\usepackage{indentfirst}

\begin{document}

\title{Quantum Foundations in the Light of \\ Quantum Information}

\author{Christopher A. Fuchs \smallskip
\\
\it Computing Science Research Center
\\
\it Bell Labs, Lucent Technologies
\\
\it Room 2C-420, 600--700 Mountain Ave.
\\
\it Murray Hill, New Jersey 07974, USA}

\date{}

\maketitle

\begin{abstract}
In this paper, I try to cause some good-natured trouble.  The issue
at stake is when will we ever stop burdening the taxpayer with
conferences and workshops devoted---explicitly or implicitly---to the
quantum foundations?  The suspicion is expressed that no end will be
in sight until a means is found to reduce quantum theory to two or
three statements of crisp {\it physical\/} (rather than abstract,
axiomatic) significance. In this regard, no tool appears to be
better calibrated for a direct assault than quantum information
theory. Far from being a strained application of the latest fad to a
deep-seated problem, this method holds promise precisely because a
large part (but not all) of the structure of quantum theory has
always concerned information.  It is just that the physics community
has somehow forgotten this.
\end{abstract}
\bigskip

\section{Imprimatur}
\begin{flushright}
\baselineskip=3pt
\parbox{3.4in}{\baselineskip=3pt\footnotesize
\bq
{\bf im$\cdot$pri$\cdot$ma$\cdot$tur}
(\^{\i}m$\!$\'{}pre-m\"{a}$^1$ter, -m\^{a}$^1$ter)\\
{\bf 1}.\ Official approval or license to print or publish, especially
under conditions of censorship.\\
\hspace*{\fill} --- {\sl American Heritage Dictionary}
\eq
}
\end{flushright}
\vspace{-.2in}

The title of the NATO Advanced Research Workshop that gave birth to
this volume was ``Decoherence and its Implications in Quantum
Computation and Information Transfer.''  It was a wonderful
meeting---the kind most of us lick our lips for year after year, with
little hope of ever tasting. It combined the best of science with
the exotic solitude of an island far, far away. One could not help
but have a creative thought shaken loose with each afternoon's gusty
wind. Indeed, it was a meeting that will make NATO proud. But, as
any attendee can tell you, the most popular pastime---in spite of
those windy beaches and dark tans---was an activity just
half-devoted to the conference title. The life of the party was all
the talks and conversations on ``Decoherence and its Implications in
Quantum {\it Foundations}.''

In this article, I am going to make an admission at the outset.
Despite the industry of work it has spawned for so many conscientious
colleagues, and the absolute importance its understanding holds for
implementing quantum information technologies, {\it I simply cannot
see that decoherence bears in any way on the quantum foundation
problem}. In saying this, I am well aware of the jeopardy I bring
upon myself. For I am but one person---one, in fact, who makes no
bones about how limited his knowledge is---while my decoherence
colleagues tell me that they are many. The press tells me that they
are the ``new orthodoxy.''   Try as I might, though, I just do not
get it. Something about their program does not click in my head.

The root of this problem could be many things, of course---not least
of which might be my supreme thickheadedness.  There is, however, one
thing I know for sure:  My want of understanding does not come from
a lack of {\it trying\/} to understand.  To the extent that I am a
relatively reasonable person---and presumably as conscientious as
the enthusiasts of the ``new orthodoxy''---it seems to me this is a
datum that should not go unnoticed. Why might it be that an honest
outsider is having so much trouble coming to grips with something
the indoctrinated profess to see with great clarity? I cannot stop
myself from thinking, ``Where there is smoke, there is smoke.''

But let me put that aside:  What I wish to get at with this
self-sanctioned imprimatur is not a detailed criticism of the
decoherence-based quantum-foundations programs.  I say all these
things instead to provide a particular instantiation (and a little
background) for what I deem to be a significantly larger problem in
the quantum-foundations efforts to date. My reluctance, or more
accurately, my inability, to toe the line for any of the rival
quantum political parties---be they the Bohmians~\cite{Cushing96},
the Consistent Historians~\cite{Griffiths99}, the
Einselectionists~\cite{Zurek98}, the Spontaneous
Collapsicans~\cite{Ghirardi91}, or the
Everettistas~\cite{Deutsch97}---springs from a distrust captured
vividly by the image of a political convention.  The relevant point
is, what are their platforms?

Throughout the 2000 presidential campaign in America, the Green Party
accused the Democrats and the Republicans of having no difference
{\it whatsoever\/} in their platforms.  The Democrats and
Republicans were appalled.  Likewise, to what I suspect will be a
jaw-dropping shock to the quantum-party leaders, I declare that I
can see little to no difference in {\it any\/} of their beliefs. They
all look equally pale to me.  For though everyone seems to want a
little reality---i.e., something in the theory that they can point
to and say, ``There, that term is what is real in the universe even
when there are no physicists about''---none are willing to dig very
deep for it.

What I mean by this deliberately provocative statement is that in
spite of the differences in what the various parties are willing to
label\footnote{Or {\it add\/} to the theory, as the case may be.}
``real'' in quantum theory,\footnote{Very briefly, a {\it cartoon\/}
of the positions might be as follows.  For the Bohmians, ``reality''
is captured by supplementing the state vector with an actual
trajectory in coordinate space.  For the Everettistas, it is the
universal wave function and the universe's Hamiltonian. (Depending
upon the faction, though, these two entities are sometimes
supplemented with the terms in various Schmidt decompositions of the
universal state vector with respect to a preconceived tensor-product
structure.) For the Spontaneous Collapsicans it is again the state
vector---though now for the individual system---but Hamiltonian
dynamics is supplemented with an objective collapse mechanism. For
the Consistent Historians ``reality'' is captured with respect to an
initial quantum state and a Hamiltonian by the addition of a set of
preferred positive-operator valued measures (POVMs)---they call them
consistent sets of histories---along with a truth-value assignment
within each of those sets.  For the Einselectionists, I leave it as
an exercise to the reader.} they nonetheless all proceed from
essentially the same {\it abstract\/} starting point. It is nothing
other than the standard textbook accounts of the {\it axioms\/} of
quantum theory.\footnote{To be fair, they do, each in their own way,
contribute minor modifications to the {\it meanings\/} of a few {\it
words\/} in the axioms.  But that is essentially where the effort
stops.}

\begin{center}
\begin{tabular}{|ll|}
\hline
& \\
\multicolumn{2}{|c|}{\it The Platform for Most Quantum %
                         Foundations Ventures:} \\
\multicolumn{2}{|c|}{\it \rm The Axioms (plain and simple)} \\
& \\
\hline\hline
& \\
\hspace{.05in}
{\bf 1.} & For every system, there is a complex Hilbert space
$\cal H$.  \\
& \\
\hspace{.05in} {\bf 2.} & States of the system correspond to
projection operators onto $\cal H$. \hspace{.05in} \\
& \\
\hspace{.05in} {\bf 3.} & Those things that are observable {\it
somehow\/}
           correspond to the \\
         & eigenprojectors 
           of Hermitian operators. \\
& \\
\hspace{.05in} {\bf 4.} & Isolated systems evolve according to the
Schr\"odinger  equation. \\
\multicolumn{2}{|c|}{\bf \vdots} \\
\hline
\end{tabular}
\end{center}

``But what nonsense is this,'' you must be asking.  ``Where else
could they start!?!''  The main issue is this, and no one has said it
more clearly than Carlo Rovelli~\cite{Rovelli96}. Where present-day
quantum-foundation studies have stagnated in the stream of history
is not so unlike where the physics of length contraction and time
dilation stood before Einstein's 1905 paper on special relativity.

The Lorentz transformations have the name that they do, rather than,
say, the Einstein transformations, for a simple historical reason:
Lorentz had published some of them as early as 1895.\footnote{Though,
FitzGerald had considered length contraction as early as 1889.}
Indeed one could say that most of the empirical predictions of
special relativity were in place well before Einstein came onto the
scene. But that was of little consolation to the pre-Einsteinian
physics community striving so hard to make sense of electromagnetic
phenomena and the luminiferous ether. Precisely because the {\it
only\/} justification for the Lorentz transformations appeared to be
their {\it empirical adequacy}, they remained a mystery to be
conquered. More particularly, this was a mystery that heaping
further {\it ad hoc\/} (mathematical) structure onto could not
possibly solve.

What was being begged for in the years between 1895 and 1905 was an
understanding of the {\it origin\/} of that abstract, mathematical
structure---some simple, crisp {\it physical\/} statements with
respect to which the necessity of the mathematics would be
indisputable. Einstein supplied that and became one of the greatest
physicists of all time.  He reduced the mysterious structure of the
Lorentz transformations to two simple statements that could be
written in any common language:
\begin{verse}
1) the speed of light in empty space is independent of the speed of
its source, \\
2) physics should appear the same in all inertial reference frames.
\end{verse}
The deep significance of this for the quantum problem should stand
out and speak overpoweringly to anyone who admits the simplicity of
these principles.

Einstein's move effectively stopped all further debate on the origins
of the Lorentz transformations.  Outside of the time of the Nazi
regime in Germany, I suspect there have been less than a handful of
conferences devoted to ``interpreting'' them. More importantly, with
the supreme simplicity of Einstein's principles, physics became
ready for ``the next step.'' Is it possible to imagine that any
mind---even Einstein's---could have made the leap to general
relativity directly from the original, abstract structure of the
Lorentz transformations?  A structure that was only empirically
adequate?  I would say no. Indeed, one might question what wonders we
will find by pursuing the same strategy of simplification for the
quantum foundations.

\begin{center}
\begin{tabular}{|c||c|} \hline
& \\
\hspace{.05in} Symbolically, where we are: \hspace{.05in} &
\hspace{.05in} Where we need to be: \hspace{.05in} \\
& \\
\hline \hline
& \\
& \\
$\displaystyle x^\prime= \frac{x - v t}{\,\sqrt{1-v^2/c^2}\,}$
& $\matrix{\mbox{Speed of light}\cr\mbox{is constant.}}
$ \\
& \\
& \\
$\displaystyle t^\prime= \frac{t - v x/c^2}{\,\sqrt{1-v^2/c^2}\,}$
& $\matrix{\mbox{Physics is the same}\cr\mbox{in all inertial frames.}}
$ \\
& \\
\hline
\end{tabular}
\end{center}

The task is not to make sense of the quantum axioms by heaping more
structure, more definitions, more science-fiction imagery on top of
them, but to throw them away wholesale and start afresh.  We should
be relentless in asking ourselves:  From what deep {\it physical\/}
principles might we {\it derive\/} this exquisite mathematical
structure?  Those principles should be crisp; they should be
compelling. They should stir the soul. When I was in junior high
school, I sat down with Martin Gardner's book {\sl Relativity for
the Million\/} and came away with an understanding of the subject
that sustains me even today:  The concepts were strange to my
everyday world, but they were clear enough that I could get a grasp
of them knowing little more mathematics than arithmetic. One should
expect nothing less for a proper foundation to the quantum. Until we
can explain the essence of the theory to a junior high-school or
high-school student---{\it the essence, not the mathematics!}---and
have them walk away with a deep, lasting memory, I well believe we
will have not understood a thing about quantum foundations.

But I am not fooling myself.  I know that anyone with a vested
interest in any of the existing quantum interpretations will be
quick to point out every hole, every nonnecessity in the sermon I
just gave. Indeed, I can feel the upcoming wrath of their email as I
write this sentence.  I have no retort.  Only a calm confidence that
if progress is not made in this direction, 100 years from now the
political pollster will still have a niche at the latest quantum
foundations meeting~\cite{Tegmark98}.

So, throw the existing axioms of quantum mechanics away and start
afresh! But how to proceed? I myself see no alternative but to
contemplate deep and hard the tasks, the techniques, and the
implications of quantum information theory. The reason is simple,
and I think inescapable.  Quantum mechanics has always been about
information.  It is just that the physics community has somehow
forgotten this.

\begin{flushleft}
\begin{tabular}{|ll||ll|}
\hline
\multicolumn{4}{|c|}{} \\
\multicolumn{4}{|c|}{\bf \small Quantum Mechanics:} \\
\multicolumn{4}{|c|}{\it \small The Axioms and Our Imperative!} \\
\multicolumn{4}{|c|}{} \\
\hline\hline
& & & \\
\hspace{.1in} \small {\bf 1.} & \small States correspond to density
& &
   \small {\it Give an information theoretic reason} \hspace{.05in} \\
   & \small operators $\rho$ over a Hilbert space $\cal H$. &  &
   \hspace{.1in} \small {\it if possible!} \\
& & & \\
\hspace{.1in} \small {\bf 2.} & \small Measurements correspond to positive & & \\
   & \small operator-valued measures (POVMs) & &
   \small {\it Give an information theoretic reason}\\
   & \small $\{E_b\}$ on $\cal H$. &  &
   \hspace{.1in} \small {\it if possible!} \\
& & & \\
\hspace{.1in} \small {\bf 3.} & \small $\cal H$ is a complex vector space, & & \\
   & \small not a real vector space, not a & &
   \small {\it Give an information theoretic reason}\\
   & \small quaternionic module. &  &
   \hspace{.1in} \small {\it if possible!} \\
& & & \\
\hspace{.1in} \small {\bf 4.} & \small Systems combine according to the tensor & & \\
   & \small product of their separate vector & &
   {\it \small Give an information theoretic reason}\\
   & \small spaces, ${\cal H}_{\rm\scriptscriptstyle AB}=
   {\cal H}_{\rm\scriptscriptstyle A}\otimes
   {\cal H}_{\rm\scriptscriptstyle B}$. &  &
   \hspace{.1in} \small {\it if possible!} \\
& & & \\
\hspace{.1in} \small {\bf 5.} & \small Between measurements, states evolve & & \\
   & \small according to trace-preserving completely & &
   \small {\it Give an information theoretic reason}\\
   & \small positive linear maps. &  &
   \hspace{.1in} \small {\it if possible!} \\
& & & \\
\hspace{.1in} \small {\bf 6.} & \small By way of measurement, states evolve & & \\
   & \small (up to normalization) via outcome- & &
   \small {\it Give an information theoretic reason}\\
   & \small dependent completely positive linear maps. \hspace{.01in} & &
   \hspace{.1in} \small {\it if possible!} \\
& & & \\
\hspace{.1in} \small {\bf 7.} & \small Probabilities for the outcomes & & \\
   & \small of a measurement obey the Born rule & &
   \small {\it Give an information theoretic reason}\\
   & \small for POVMs ${\rm tr}(\rho E_b)$. & &
   \hspace{.1in} {\it \small if possible!} \\
& & & \\
\hline
\end{tabular}
\end{flushleft}

This table is my plea to the community. Our foremost task should be
to go to each and every axiom of quantum theory and give it an
information theoretic justification if we can. Only when we are
finished picking off all the terms (or combinations of terms) that
can be interpreted as information---subjective information---will we
be in a position to make real progress. The raw distillate that is
left behind, miniscule though it may be, will be our first glimpse
of what quantum mechanics is trying to tell us about nature itself.

\section{Introduction}

This paper is about taking that plea to heart, though it contributes
only a small amount to the labor it asks.  Just as in the founding
of quantum mechanics, this is not something that will spring forth
from a lone mind in the shelter of a medieval college. It is a task
for a community with diverse but productive points of view. The
quantum information community is nothing if not that. ``Philosophy
is too important to be left to the philosophers,'' John Archibald
Wheeler once said. Likewise, I am apt to say the same for the
quantum foundations.

The structure of the remainder of the paper is as follows.  In
Section 3 ``{\bf Why Information?},'' I reiterate the cleanest
argument I know of that the quantum state is solely an expression of
information---the information one has about a quantum system.  It
has no objective reality in and of itself.  The argument is then
refined by considering the phenomenon of quantum
teleportation~\cite{Bennett93}.

In Section 4 ``{\bf Information About What?},'' I tackle that very
question~\cite{Bub00} head-on.  The answer is, ``nothing more than
the potential consequences of our experimental interventions into
nature.'' Once freed from the notion that quantum measurement ought
to be about revealing traces of some preexisting
property~\cite{Bub97} (or beable \cite{Bell87}), one finds no
particular reason to take the standard account of measurement (in
terms of complete sets of orthogonal projection operators) as a
basic notion. Indeed quantum information theory, with its emphasis
on the utility of generalized measurements or positive
operator-valued measures (POVMs)~\cite{Nielsen00}, suggests one
should take those entities as the basic notion instead.  The
productivity of this point of view is demonstrated by the
beautifully simple Gleason-like derivation of the quantum
probability rule recently found by Paul Busch~\cite{Busch99} and,
independently, by Joseph Renes and collaborators~\cite{Renes00}.
Contrary to Gleason's original theorem~\cite{Gleason57}, this theorem
works just as well for two-dimensional Hilbert spaces, and even for
Hilbert spaces over the field of rational numbers.

In Section~5 ``{\bf Whither Bayes Rule?},'' I ask why one should
expect the rule for updating quantum state assignments upon the
completion of a measurement to take the form it actually does. Along
the way, I give a simple derivation that one's information always
increases on average for {\it any\/} quantum mechanical measurement
that does not itself discard information.  (Despite the appearance
otherwise, this is not a tautology!) More importantly, the proof
technique used for showing the theorem indicates an extremely strong
analogy between quantum collapse and Bayes' rule in classical
probability theory: Up to an overall unitary ``readjustment'' of
one's knowledge (that takes into account details of the measurement
interaction as well as one's initial state of knowledge), quantum
collapse is {\it precisely\/} Bayesian conditionalization. This in
turn gives even more impetus for the assumptions behind the
Gleason-like theorem of the previous section.

In Section 6 ``{\bf Wither Entanglement?},'' I ask whether
entanglement is all it is cracked-up to be as far as quantum
foundations are concerned.  In particular, I give a simple
derivation of the tensor-product rule for combining Hilbert spaces
of individual systems into a larger composite system.  To no
surprise, once again this comes from Gleason-like considerations for
local measurements in the presence of classical communication---the
very bread and butter of much of quantum information theory.

In Section 7 ``{\bf Unknown Quantum States?},'' I tackle the
conundrum posed by these very words.  Despite the phrase's
ubiquitous use in the quantum information literature, what can an
{\it unknown\/} state possibly be?  A quantum state---from the
present point of view, explicitly someone's information---must always
be known by someone, if it exists at all.  On the other hand, for
many an application in quantum information, it would be quite a
contrivance to imagine there is always someone in the background
with extra knowledge of the system being measured or manipulated.
The solution, at least in the case of quantum state
tomography~\cite{Vogel89}, is found through a quantum mechanical
version of de Finetti's classic theorem on ``unknown
probabilities.''  This reports work from Refs.~\cite{Caves01} and
\cite{Schack00}. Maybe one of the most interesting things about the
theorem is that it fails for Hilbert spaces over the field of real
numbers, suggesting that perhaps the whole discipline of quantum
information might not be well defined in that imaginary world.

Finally, in Section 8 ``{\bf The Oyster and the Quantum},'' I flirt
with the most tantalizing question of all:  Why the quantum? There
are no answers here, but I do not discount that there will be one
within 20 years. In this regard no platform seems firmer for the
leap than the very existence of quantum cryptography and quantum
computing. The world is sensitive to our touch.  It has a kind of
``Zing!''\footnote{Dash, verve, vigor, vim, zip, pep, punch,
pizzazz!}~that makes it fly off in ways that were not imaginable
classically.  The whole structure of quantum mechanics---{\it
\underline{it is speculated}}---may be nothing more than the optimal
method of reasoning and processing information in the light of such
a fundamental (wonderful) sensitivity.

\section{Why Information?}
\begin{flushright}
\parbox{3.4in}{\footnotesize
\bq
It may be, as one French physicist put it, ``the fog from the north,''
but the Copenhagen interpretation remains the best interpretation of
the quantum that we \medskip have.
\hspace*{\fill} --- {\it John Archibald Wheeler}\\
\hspace*{\fill} \footnotesize New York Times, 12 December 2000
\eq
}
\end{flushright}

Einstein was the master of clear thought. I have already expressed my
reasons for thinking this in the arena of electromagnetic phenomena.
Likewise, I would say he possessed the same great penetrating power
when it came to analyzing the quantum. For even there, he was
immaculately clear and concise in his expression. In particular, he
was the first person to say in absolutely unambiguous terms why the
quantum state should be viewed as information (or, to say the same
thing, as a representation of one's knowledge).

His argument was simply that a quantum-state assignment for a system
can be forced to go one way or the other by interacting with a part
of the world that should have no causal connection with the system
of interest.  The paradigm here is of course the one well known
through the Einstein, Podolsky, Rosen paper~\cite{Einstein35}, but
simpler versions of the train of thought had a long pre-history with
Einstein~\cite{Fine86}.

The best was in essence this.  Take two spatially separated systems
$A$ and $B$ prepared in some entangled quantum state
$|\psi^{\scriptscriptstyle AB}\rangle$. By performing the
measurement of one or another of two observables on system $A$
alone, one can {\it immediately\/} write down a new state for system
$B$.  Either the state will be drawn from one set of states
$\{|\phi_i^{\scriptscriptstyle B}\rangle\}$ or another
$\{|\eta_i^{\scriptscriptstyle B}\rangle\}$, depending upon which
observable is measured.\footnote{Generally there need be hardly any
relation between the two sets of states:  only that when the states
are weighted by their probabilities, they mix together to form the
initial density operator for system $B$ alone. For a precise
statement of this freedom, see Ref.~\cite{Hughston93}.} The key
point is that it does not matter how distant the two systems are
from each other, what sort of medium they might be immersed in, or
any of the other fine details of the world.  Einstein concluded that
whatever these things called quantum states {\it be}, they cannot be
``real states of affairs'' for system $B$ alone.  For, whatever the
real, objective state of affairs at $B$ is, it should not depend
upon the measurements one can make on a causally unconnected system
$A$.

Thus one must take it seriously that the new state (either a
$|\phi_i^{\scriptscriptstyle B}\rangle$ or a
$|\eta_i^{\scriptscriptstyle B}\rangle$) represents partial
knowledge about system $B$.  In making a measurement on $A$, one
learns something about $B$, but that is where the story ends. The
state change cannot be construed to be something more physical than
that. More particularly, the state itself cannot be viewed as more
than a reflection of the knowledge gained through the measurement.
Expressed in the language of Einstein, the quantum state cannot be a
``complete'' description of the quantum system.

Here is the way Einstein put it to Michele Besso in a 1952
letter~\cite{Bernstein91}:
\begin{quotation}
\small What relation is there between the ``state'' (``quantum
state'') described by a function $\psi$ and a real deterministic
situation (that we call the ``real state'')?  Does the quantum state
characterize completely (1) or only incompletely (2) a real state?
\ldots

I reject (1) because it obliges us to admit that there is a rigid
connection between parts of the system separated from each other in
space in an arbitrary way (instantaneous action at a distance, which
doesn't diminish when the distance increases).  Here is the
demonstration: \ldots

If one considers the method of the present quantum theory as being in
principle definitive, that amounts to renouncing a complete
description of real states.  One could justify this renunciation if
one assumes that there is no law for real states---i.e., that their
description would be useless.  Otherwise said, that would mean: laws
don't apply to things, but only to what observation teaches us about
them.  (The laws that relate to the temporal succession of this
partial knowledge are however entirely deterministic.)

Now, I can't accept that.  I think that the statistical character of
the present theory is simply conditioned by the choice of an
incomplete description.
\end{quotation}

There are two issues in this letter that are worth disentangling. 1)
Rejecting the rigid connection of all nature\footnote{The rigid
connection of all nature, on the other hand, is exactly what the
Bohmians and Everettistas {\it do} embrace, even glorify.  So, I
suspect these words will fall on deaf ears with them.  But similarly
would they fall on deaf ears with the believer who says that God
wills each and every event in the universe and no further
explanation is needed. No point of view should be dismissed out of
hand:  the overriding issue is simply which view will lead to the
most progress, which view has the potential to close the debate,
which view will give the most new phenomena for the physicist to
have fun with?}---that is to say, admitting that the very notion of
{\it separate systems\/} has any meaning at all---one is led to the
conclusion that a quantum state cannot be a complete specification
of a system.  It must be information, at least in part.  This point
should be placed in contrast to the other well-known facet of
Einstein's thought: namely, 2) an unwillingness to accept such an
``incompleteness'' as a necessary trait of the physical world.

It is quite important to recognize that the first issue does not
entail the second.  Einstein had that firmly in mind, but he wanted
more. His reason for going the further step was, I think, well
justified {\it at the time\/}~\cite{Einstein70}:
\begin{quotation}
\small There exists [...]\ a simple psychological reason for the fact
that this most nearly obvious interpretation is being shunned.  For
if the statistical quantum theory does not pretend to describe the
individual system (and its development in time) completely, it
appears unavoidable to look elsewhere for a complete description of
the individual system; in doing so it would be clear from the very
beginning that the elements of such a description are not contained
within the conceptual scheme of the statistical quantum theory.  With
this one would admit that, in principle, this scheme could not serve
as the basis of theoretical physics.
\end{quotation}

But the world has seen much in the mean time.  The last seventeen
years have given confirmation after confirmation that the Bell
inequality (and several variations of it) are indeed violated by the
physical world.  The Kochen-Specker no-go theorems have been
meticulously clarified to the point where simple textbook pictures
can be drawn of them~\cite{Peres93}.  Incompleteness, it seems, is
here to stay:  The theory prescribes that no matter how much we know
about a quantum system---even when we have {\it maximal\/}
information about it---there will always be a statistical residue.
There will always be questions that we can ask of a system for which
we cannot predict the outcomes.  In quantum theory, maximal
information is simply not complete information~\cite{Caves96}.  But
neither can it be completed.  As Wolfgang Pauli once wrote to Markus
Fierz~\cite{Pauli54}, ``The well-known `incompleteness' of quantum
mechanics (Einstein) is certainly an existent fact
somehow-somewhere, but certainly cannot be removed by reverting to
classical field physics.''  Nor, I would add, will the mystery of
that ``existent fact'' be removed by attempting to give the quantum
state anything resembling an ontological status.

The complete disconnectedness of the quantum-state change rule from
anything to do with spacetime considerations is telling us something
deep: The quantum state is information. Subjective, incomplete
information. Put in the right mindset, this is {\it not\/} so
intolerable.  It is a statement about our world. There is something
about the world that keeps us from ever getting more information
than can be captured through the formal structure of quantum
mechanics. Einstein had wanted us to look further---to find out how
the incomplete information could be completed---but perhaps the real
question is, ``Why can it {\it not\/} be completed?''

Indeed I think this is one of the deepest questions we can ask and
still hope to answer. But first things first. The more immediate
question for anyone who has come this far---and one that deserves to
be answered forthright---is what is this information symbolized by a
$|\psi\rangle$ actually about? I have hinted that I would not dare
say that it is about some kind of hidden variable (as the Bohmian
might) or even about our place within the universal wavefunction (as
the Everettista might).

Perhaps the best way to build up to an answer is to be true to the
title of this paper:  quantum foundations in the light of quantum
information.  Let us forage the phenomena of quantum information to
see if we might first refine Einstein's argument.  One need look no
further than to the phenomenon of quantum
teleportation~\cite{Bennett93}.  Not only can a quantum-state
assignment for a system be forced to go one way or the other by
interacting with another part of the world of no causal significance,
but, for the cost of two bits, one can make that quantum state
assignment anything one wants it to be.

Such an experiment starts out with Alice and Bob sharing a maximally
entangled pair of qubits in the state $|\psi^{\scriptscriptstyle
AB}\rangle=\sqrt{\frac{1}{2}}\,(|0\rangle|0\rangle+|1\rangle|1\rangle)$.
Bob then goes to any place in the universe he wishes. Alice in her
laboratory prepares another qubit with any state $|\psi\rangle$ that
she ultimately wants to impart onto Bob's system.  She performs a
Bell-basis measurement on the two qubits in her possession.  In the
same vein as Einstein's thought experiment, Bob's system immediately
takes on the character of one of the states $|\psi\rangle$, $\sigma_x
|\psi\rangle$, $\sigma_y |\psi\rangle$, or $\sigma_z |\psi\rangle$.
But that is only insofar as Alice is concerned.\footnote{As far as
Bob is concerned, nothing whatsoever changes about the system in his
possession:  It started in the completely mixed state $\rho =
\frac{1}{2}I$ and remains that way.} Since there is no (reasonable)
causal connection between Alice and Bob, it must be that these
states represent the possibilities for Alice's new knowledge of Bob's
system.

If now Alice broadcasts the result of her measurement to the world,
Bob may complete the teleportation protocol by performing one of the
four Pauli rotations ($I$, $\sigma_x$, $\sigma_y$, $\sigma_z$) on
his system, conditioning it on the information he receives.  The
result, as far as Alice is concerned, is that Bob's system finally
resides predictably in the state $|\psi\rangle$.\footnote{As far as
Bob is concerned, nothing whatsoever changes about the system in his
possession:  It started in the completely mixed state $\rho =
\frac{1}{2}I$ and remains that way.}

How can Alice convince herself that such is the case?  Well, if Bob
is willing to reveal his location, she just need walk to his site and
perform the {\bf YES}-{\bf NO} measurement:
$|\psi\rangle\langle\psi|$ vs.\ $I-|\psi\rangle\langle\psi|$.  The
outcome will be a {\bf YES} with probability one if all has gone well
in carrying out the protocol.  Thus, for the cost of a measurement on
a causally disconnected system and two bits worth of causal action
on the system of actual interest---i.e., one of the four Pauli
rotations---Alice can sharpen her predictability to complete
certainty for {\it any\/} {\bf YES}-{\bf NO} observable she wishes.

Roger Penrose argues in his book {\sl The Emperor's New Mind\/}
\cite{Penrose89} that when a system ``has'' a state $|\psi\rangle$
there ought to be some property in the system (in and of itself) that
corresponds to its ``$|\psi\rangle$'ness.'' For how else could the
system be prepared to reveal a {\bf YES} in the case that Alice
actually checks it?  Asking this rhetorical question with a
sufficient amount of command is enough to make many a would-be
informationist weak in knees.  But there is a crucial oversight
implicit in its confidence, and we have already caught it in
action.  If Alice fails to reveal her information to anyone else in
the world, there is no one else who can predict the qubit's ultimate
revelation with certainty.  More importantly, there is nothing in
quantum mechanics that gives the qubit the power to stand up and say
{\bf YES} all by itself:  If Alice does not take the time to walk
over to it and interact with it, there is no revelation. There is
only the confidence in Alice's mind that, {\it should\/} she
interact with it, she {\it could\/} predict the consequence of that
interaction.

\section{Information About What?}
\begin{center}
\begin{tabular}{ll}
\parbox{3.4in}{\footnotesize
\bq
[S]urely, the existence of [the] world is the primary experimental
fact of all, without which there would be no point to physics or any
other science; and for which we all receive new evidence every waking
minute of our lives.  This direct evidence of our senses is vastly
more cogent than are any of the deviously indirect experiments that
are cited as evidence for the Copenhagen interpretation.
\\
\hspace*{\fill} --- {\it E.~T. Jaynes, 1986}
\eq
}
&
\parbox{3.4in}{\footnotesize
\bq
The criticism of the Copenhagen interpretation of the quantum theory
rests quite generally on the anxiety that, with this interpretation,
the concept of ``objective reality'' which forms the basis of
classical physics might be driven out of physics. \ldots\ [T]his
anxiety is groundless \ldots\  At this point we realize the simple fact
that natural science is not Nature itself but a part of the relation
between Man and Nature, and therefore is dependent on Man.
\\
\hspace*{\fill} --- {\it Werner Heisenberg, 1955}
\eq
}
\end{tabular}
\end{center}

There are great rewards in being a new parent.  Not least of all is
the opportunity to have a close-up look at a mind in formation. I
have been watching my two year old daughter learn things at a
fantastic rate, and though there have been untold numbers of lessons
for her, there have also been a sprinkling for me.  For instance, I
am just starting to see her come to grips with the idea that there
is a world independent of her desires.  What strikes me is the
contrast between this and the concomitant gain in confidence I see
grow in her everyday that there are aspects of existence she
actually {\it can\/} control.  The two go hand in hand.  She pushes
on the world, and sometimes it gives in a way that she has learned
to predict, and sometimes it pushes back in a way she has not
foreseen (and may never be able to). If she could manipulate the
world to the complete desires of her will, I am quite sure, there
would be little difference between wake and dream.

But the main point is that she learns from her forays into the
world.  In my more cynical moments, I find myself thinking, ``How
can she think that she's learned anything at all?  She has no theory
of measurement.  She leaves measurement completely undefined.  How
can she have any true stake to knowledge?''

Hideo Mabuchi once told me, ``The quantum measurement problem refers
to a set of people.''  And though that is a bit harsh, maybe it also
contains a bit of the truth.  With the physics community making use
of theories that tend to last between 100 and 300 years, we are apt
to forget that scientific views of the world are built from the top
down, not from the bottom up.   The experiment is the basis of all
that we know to be firm.  But an experiment is an active intervention
into the course of nature on the part of the experimenter; it is not
contemplation of nature from afar~\cite{Jammer84}. We set up this or
that experiment to see how nature reacts. It is the conjunction of
myriads of such interventions and their consequences that we record
into our data books.\footnote{But I must stress that I am not so
positivistic as to think that physics should somehow be grounded on
a primitive notion of ``sense impression'' as the philosophers of the
Vienna Circle did.  The interventions and their consequences that an
experimenter records, have no option but to be thoroughly
theory-laden.  It is just that, in a sense, they are by necessity at
least one theory behind.  No one got closer to the salient point than
Heisenberg (in a quote he attributed to Einstein many years after the
fact)~\cite{Heisenberg71}:
\begin{quote}
It is quite wrong to try founding a theory on observable magnitudes
alone.  In reality the very opposite happens.  It is the theory which
decides what we can observe.  You must appreciate that observation is
a very complicated process.  The phenomenon under observation
produces certain events in our measuring apparatus.  As a result,
further processes take place in the apparatus, which eventually and
by complicated paths produce sense impressions and help us to fix the
effects in our consciousness.  Along this whole path---from the
phenomenon to its fixation in our consciousness---we must be able to
tell how nature functions, must know the natural laws at least in
practical terms, before we can claim to have observed anything at
all.  Only theory, that is, knowledge of natural laws, enables us to
deduce the underlying phenomena from our sense impressions.  When we
claim that we can observe something new, we ought really to be saying
that, although we are about to formulate new natural laws that do not
agree with the old ones, we nevertheless assume that the existing
laws---covering the whole path from the phenomenon to our
consciousness---function in such a way that we can rely upon them and
hence speak of ``observation.''
\end{quote}\vspace{-.15in}}

We tell ourselves that we have learned something new when we can
distill from the data a compact description of all that was seen
and---even more tellingly---when we can dream up further experiments
to corroborate that description. This is the minimal requirement of
science.  If, however, from such a description we can {\it
further\/} distill a model of a free-standing ``reality''
independent of our interventions, then so much the better.  I have
no bone to pick with reality.  It is the most solid thing we can
hope for from a theory.\footnote{Woody Allen said it best: ``I hate
reality, but, you know, where else can you get a good steak
dinner?'' Spending my childhood in Texas, where beef is the staple
of most meals, the reader should realize that it has been no easy
journey for me to come to my present view of quantum mechanics!}
Classical physics is the ultimate example in that regard. It gives
us a compact description, but it can give much more if we want it to.

The important thing to realize, however, is that there is no logical
necessity that such a worldview always be obtainable.  If the world
is such that we can never identify a reality---a {\it
free-standing\/} reality---independent of our experimental
interventions, then we must be prepared for that too. That is where
quantum theory in its most minimal and conceptually simplest
dispensation seems to stand~\cite{Fuchs00}. It is a theory whose
terms refer predominately to our interface with the world.  It is a
theory that cannot go the extra step that classical physics did
without ``writing songs I can't believe, with words that tear and
strain to rhyme''~\cite{Simon65}. It is a theory not about
observables, not about beables, but about ``dingables.''  We tap the
bell with our gentle touch and listen for its beautiful ring.

So what are the ways we can intervene on the world?  What are the
ways we can push it and wait for its unpredictable reaction?  The
usual textbook story is that those things that are measurable
correspond to Hermitian operators.  Or perhaps to say it in more
modern language, to each observable there corresponds a set of
orthogonal projection operators $\{\Pi_i\}$ over a complex Hilbert
space ${\cal H}_d$ that form a complete resolution of the identity,
\be
\sum_i \Pi_i = I\;.
\ee
The index $i$ labels the potential outcomes of the measurement (or
{\it intervention\/}, to slip back into the language promoted above).
When an observer possesses a state of knowledge $\rho$---captured
most generally by a mixed-state density operator---quantum mechanics
dictates that he can expect the various outcomes with a probability
\be
P(i)=\tr (\rho \Pi_i)\;.
\ee

The best justification for this probability rule comes by way of
Andrew Gleason's amazing 1957 theorem~\cite{Gleason57}.  For, it
states that the standard rule is the {\it only\/} rule that
satisfies a very simple kind of noncontextuality for measurement
outcomes~\cite{Barnum00}. In particular, if one contemplates
measuring two distinct observables $\{\Pi_i\}$ and
$\{\tilde{\Pi}_i\}$ which happen to share a single projector
$\Pi_k$, then the probability of outcome $k$ is independent of which
observable it is associated with.  More formally, the statement is
this.  Let ${\cal P}_d$ be the set of projectors associated with a
(real or complex) Hilbert space ${\cal H}_d$ for $d\ge3$, and let
$f:{\cal P}_d\longrightarrow[0,1]$ be such that
\be
\sum_i f(\Pi_i) = 1
\ee
whenever a set of projectors $\{\Pi_i\}$ forms an observable. The
theorem concludes that there exists a density operator $\rho$ such
that
\be
f(\Pi)=\tr (\rho \Pi)\;.
\ee
In fact, in a single blow, Gleason's theorem derives not only the
probability rule, but also the state-space structure for quantum
mechanical states (i.e., that it corresponds to the convex set of
density operators).

In itself this is no small feat, but the thing that makes the theorem
an ``amazing'' theorem is the shear difficulty required to prove
it~\cite{Cooke81}. Note that no restrictions have been placed upon
the function $f$ beyond the ones mentioned above. There is no
assumption that it need be differentiable, nor that it even need be
continuous.  All of that, and linearity too, comes from the
structure of the observables---i.e., that they are complete sets of
orthogonal projectors onto a linear vector space.

Nonetheless, one should ask:  Does this theorem really give {\it the
physicist\/} a clearer vision of where the probability rule comes
from? Astounding feats of mathematics are one thing; insight into
physics is another. The two are often at opposite ends of the
spectrum. As fortunes turn however, a unifying strand can indeed be
drawn by viewing quantum foundations in the light of quantum
information.

The place to start is to drop the fixation that the basic set of
observables in quantum mechanics are complete sets of orthogonal
projectors.  In quantum information theory it has been found to be
extremely convenient to expand the notion of measurement to also
include general positive operator-valued measures
(POVMs)~\cite{Peres93,Kraus83}.  In other words, in place of the
usual textbook notion of measurement, {\it any\/} set $\{E_b\}$ of
positive-semidefinite operators on ${\cal H}_d$ that forms a
resolution of the identity, i.e., that satisfies
\begin{equation}
\langle\psi|E_b|\psi\rangle\ge0\,,\quad\mbox{for all
$|\psi\rangle\in{\cal H}_d$}
\label{Hank}
\end{equation}
and
\begin{equation}
\sum_b E_b = I\;,
\label{Hannibal}
\end{equation}
counts as a measurement. The outcomes of the measurement are
identified with the indices $b$, and the probabilities of the
outcomes are computed according to a generalized Born rule,
\begin{equation}
P(b)=\tr\big(\rho E_b\big) \;.
\end{equation}
The set $\{E_b\}$ is called a POVM, and the operators $E_b$ are
called POVM elements. (In the nonstandard language promoted earlier,
the set $\{E_b\}$ signifies an intervention into nature, while the
individual $E_b$ represent the potential consequences of that
intervention.) Unlike standard measurements, there is no limitation
on the number of values the index $b$ can take. Moreover, the $E_b$
may be of any rank, and there is no requirement that they be
mutually orthogonal.

The way this expansion of the notion of measurement is {\it
usually\/} justified is that any POVM can be represented formally as
a standard measurement on an ancillary system that has interacted in
the past with the system of actual interest. Indeed, suppose the
system and ancilla are initially described by the density operators
$\rho_{\scriptscriptstyle\rm S}$ and $\rho_{\scriptscriptstyle\rm A}$
respectively. The conjunction of the two systems is then described
by the initial quantum state
\begin{equation}
\rho_{\scriptscriptstyle\rm SA}=\rho_{\scriptscriptstyle\rm
S}\otimes\rho_{\scriptscriptstyle\rm A}\;.
\label{MushMush}
\end{equation}
An interaction between the systems via some unitary time evolution
leads to a new state
\begin{equation}
\rho_{\scriptscriptstyle\rm SA}\;\longrightarrow\;
U\rho_{\scriptscriptstyle\rm SA} U^\dagger\;.
\label{Aluminium}
\end{equation}
Now, imagine a standard measurement on the ancilla.  It is described
on the total Hilbert space via a set of orthogonal projection
operators $\{I\otimes\Pi_b\}$. An outcome $b$ will be found, by the
standard Born rule, with probability
\begin{equation}
P(b)={\rm tr}\!\left( U(\rho_{\scriptscriptstyle\rm
S}\otimes\rho_{\scriptscriptstyle\rm A})
U^\dagger(I\otimes\Pi_b)\right)\;.
\label{RibTie}
\end{equation}
The number of outcomes in this seemingly indirect notion of
measurement is limited only by the dimensionality of the ancilla's
Hilbert space---in principle, there can be arbitrarily many.

As advertised, it turns out that the probability formula above can
be expressed in terms of operators on the system's Hilbert space
alone: This is the origin of the POVM. If we let $|s_\alpha\rangle$
and $|a_c\rangle$ be an orthonormal basis for the system and ancilla
respectively, then $|s_\alpha\rangle|a_c\rangle$ will be a basis for
the composite system. Using the cyclic property of the trace in
Eq.~(\ref{RibTie}), we get
\bea
P(b)
&=&
\sum_{\alpha c}\langle s_\alpha|\langle a_c|\Big( (\rho_{\rm
s}\otimes\rho_{\scriptscriptstyle\rm A})  U^\dagger(I\otimes\Pi_b)
U\Big)|s_\alpha \rangle|a_c\rangle
\nonumber\\
&=&
\sum_\alpha\langle s_\alpha|\,\rho_{\scriptscriptstyle\rm
S}\!\left(\sum_c\langle a_c|\Big(
(I\otimes\rho_{\scriptscriptstyle\rm A}) U^\dagger(I \otimes\Pi_b)
U\Big)| a_c\rangle\!\right)\!|s_\alpha\rangle\;. \rule{0mm}{8mm}
\eea
Letting ${\rm tr}_{\scriptscriptstyle\rm A}$ and ${\rm
tr}_{\scriptscriptstyle\rm S}$ denote partial traces over the system
and ancilla, respectively, it follows that
\be
P(b)={\rm tr}_{\scriptscriptstyle\rm S}(\rho_{\scriptscriptstyle\rm
S} E_b)\;,
\ee
where
\be
E_b={\rm tr}_{\scriptscriptstyle\rm A}\!\left(
(I\otimes\rho_{\scriptscriptstyle\rm A}) U(I\otimes\Pi_b)
U^\dagger\right)
\label{oncogene}
\ee
is an operator acting on the Hilbert space of the original system.
This proves half of what is needed, but it is also straightforward
to go in the reverse direction---i.e., to show that for any POVM
$\{E_b\}$, one can pick an ancilla and find operators
$\rho_{\scriptscriptstyle\rm A}$, $U$, and $\Pi_b$ such that
Eq.~(\ref{oncogene}) is true.

Putting this all together, there is a sense in which standard
measurements capture everything that can be said about quantum
measurement theory~\cite{Kraus83}.  As became clear above, a way to
think about this is that by learning something about the ancillary
system through a standard measurement, one in turn learns something
about the system of real interest. Indirect though it may seem, this
can be a powerful technique, sometimes revealing information that
could not have been revealed otherwise~\cite{Holevo73}.  A very
simple example is where a sender has only a single qubit available
for the sending one of three potential messages. She therefore has a
need to encode the message in one of three preparations of the
system, even though the system is a two-state system. To recover as
much information as possible, the receiver might (just intuitively)
like to perform a measurement with three distinct outcomes.  If,
however, he were limited to a standard quantum measurement, he would
only be able to obtain two outcomes. This---perhaps
surprisingly---generally degrades his opportunities for recovery.

What I would like to bring up is whether this standard way of
justifying the POVM is the most productive point of view one can
take.  Might any of the mysteries of quantum mechanics be alleviated
by taking the POVM as a basic notion of measurement?  Does the
POVM's utility portend a larger role for it in the foundations of
quantum mechanics?

\begin{center}
\begin{tabular}{|c||c|}
\hline
& \\
\it Standard & \it Generalized
\\
\it Measurements & \it Measurements
\\
& \\
\hline\hline
& \\
$\{\Pi_i\}$ & $\{E_b\}$
\\
& \\
\hspace{.1in} $\langle\psi|\Pi_i|\psi\rangle\ge0\,,\;\forall
|\psi\rangle$ \hspace{.1in}
& \hspace{.1in} $\langle\psi|E_b|\psi\rangle\ge0\,,\;\forall
|\psi\rangle$ \hspace{.1in}
\\
& \\
$\sum_i \Pi_i = I$ & $\sum_b E_b = I$
\\
& \\
$P(i)=\tr (\rho \Pi_i)$ & $P(b)=\tr (\rho E_b)$
\\
& \\
$\Pi_i\Pi_j=\delta_{ij}\,\Pi_i$ & {\bf ---------}
\\
& \\
\hline
\end{tabular}
\end{center}

I try to make the point dramatic in my lectures by exhibiting a
transparency of the table above.  On the left-hand side there is a
list of various properties for the standard notion of a quantum
measurement. On the right-hand side, there is an almost identical
list of properties for the POVMs.  The only difference between the
two columns is that the right-hand one is {\it missing\/} the
orthonormality condition required of a standard measurement. The
question I ask the audience is this:  Does the addition of that one
extra assumption really make the process of measurement any less
mysterious?  Indeed, I imagine myself teaching quantum mechanics for
the first time and taking a vote with the best audience of all, the
students.\footnote{I am making the safe bet that they will be lucky
enough to not yet be conditioned by years of squabbles in quantum
foundations.} ``Which set of postulates for quantum measurement
would you prefer?'' I am quite sure they would respond with a blank
stare. But that is the point! It would make no difference to them,
and it should make no difference to us. The only issue worth
debating is which notion of measurement will allow us to see more
deeply into quantum mechanics.

Therefore let us pose the question that Gleason did, but with POVMs.
In other words, let us suppose that the sum total of ways an
experimenter can intervene on a quantum system corresponds to the
full set of POVMs on its Hilbert space ${\cal H}_d$.  It is the task
of the theory to give him probabilities for the various consequences
of his interventions.  Concerning those probabilities, let us (in
analogy to Gleason) assume only that whatever the probability for a
given consequence $E_c$ is, it does not depend upon whether $E_c$ is
associated with the POVM $\{E_b\}$ or, instead, any other one
$\{\tilde{E}_b\}$. This means we can assume there exists a function
\be
f:{\cal E}_d\longrightarrow[0,1]\;,
\label{Yaniggle}
\ee
where
\be
{\cal E}_d=\Big\{E:\,0\le\langle\psi|E|\psi\rangle\le
1\;,\;\forall\;|\psi\rangle\in{\cal H}_d\Big\}\;,
\ee
such that whenever $\{E_b\}$ forms a POVM,
\be
\sum_b f(E_b) = 1\;.
\ee
(In general, we will call any function satisfying $f(E)\ge0$ and
$\sum_b f(E_b) = \mbox{const}$.\ a {\it frame function}, in analogy
to Gleason's nonnegative frame functions.)

It will come as no surprise, of course, that a Gleason-like theorem
must hold for the function in Eq.~(\ref{Yaniggle}). Namely, it can
be shown that there must exist a density operator $\rho$ for which
\be
f(E)=\tr (\rho E)\;.
\ee
This was recently shown by Paul Busch~\cite{Busch99} and,
independently, by Joseph Renes and collaborators~\cite{Renes00}.
What {\it is\/} surprising however is the utter simplicity of the
proof.  Let us exhibit the whole thing right here and now.
\bq
\small
\indent First, consider the case where ${\cal H}_d$ and the operators
on it are defined {\it only\/} over the field of (complex) rational
numbers. It is no problem to see that $f$ is ``linear'' with respect
to positive combinations of operators that never go outside ${\cal
E}_d$.  For consider a three-element POVM $\{E_1,E_2,E_3\}$. By
assumption $f(E_1)+f(E_2)+f(E_3)=1$. However, we can also group the
first two elements in this POVM to obtain a new POVM, and must
therefore have $f(E_1+E_2)+f(E_3)=1$. In other words, the function
$f$ must be additive with respect to a fine-graining operation:
\be
f(E_1+E_2)= f(E_1)+f(E_2)\;.
\ee
Similarly for any two integers $m$ and $n$,
\be
f(E)=m\, f\!\left(\frac{1}{m}E\right) = n\,
f\!\left(\frac{1}{n}E\right)
\ee
Suppose $\frac{n}{m}\le1$.  Then if we write $E=nG$, this statement
becomes:
\be
f\!\left(\frac{n}{m}G\right)=\frac{n}{m}f(G)\;.
\ee
Thus we immediately have a kind of limited linearity on ${\cal E}_d$.

One {\it might\/} imagine using this property to cap off the theorem
in the following way.  Clearly the full $d^2$-dimensional vector
space ${\cal O}_d$ of Hermitian operators on ${\cal H}_d$ is spanned
by the set ${\cal E}_d$ since that set contains, among other things,
all the projection operators. Thus, we can write any operator
$E\in{\cal E}_d$ as a linear combination
\be
E=\sum_{i=1}^{d^2} \alpha_i E_i
\label{Hortense}
\ee
for some fixed operator-basis $\{E_i\}_{i=1}^{d^2}$.  ``Linearity''
of $f$ would then give
\be
f(E)=\sum_{i=1}^{d^2} \alpha_i f(E_i)\;.
\ee
So, if we define $\rho$ by solving the $d^2$ linear equations
\be
\tr(\rho E_i)=f(E_i)\;,
\ee
we would have
\be
f(E)=\sum_i \alpha_i \tr\big(\rho E_i\big) = \tr\!\left(\rho\sum_i
\alpha_i E_i\right)= \tr(\rho E)
\label{HollyHobie}
\ee
and essentially be done.  (Positivity and normalization of $f$ would
require $\rho$ to be an actual density operator.)  But the {\it
problem\/} is that in expansion~(\ref{Hortense}) there is no
guarantee that the coefficients $\alpha_i$ can be chosen so that
$\alpha_i E_i\in{\cal E}_d$.

What remains to be shown is that $f$ can be extended uniquely to a
function that is truly linear on ${\cal O}_d$.  This too is rather
simple.  First, take any positive semi-definite operator $E$. We can
always find a positive rational number $g$ such that $E=gG$ and
$G\in{\cal E}_d$.  Therefore, we can simply define $f(E)\equiv g
f(G)$.  To see that this definition is unique, suppose there are two
such operators $G_1$ and $G_2$ (with corresponding numbers $g_1$ and
$g_2$) such that $E=g_1G_1=g_2G_2$.  Further suppose $g_2\ge g_1$.
Then $G_2=\frac{g_1}{g_2}G_1$ and, by the homogeneity of the original
unextended definition of $f$, we obtain $g_2 f(G_2) = g_1 f(G_1)$.
Furthermore this extension retains the additivity of the original
function.  For suppose that neither $E$ nor $G$, though positive
semi-definite, are necessarily in ${\cal E}_d$.  We can find a
positive rational number $c\ge1$ such that $\frac{1}{c}(E+G)$,
$\frac{1}{c}E$, and $\frac{1}{c}G$ are all in ${\cal E}_d$.  Then,
by the rules we have already obtained,
\be
f(E+G)=c\, f\!\left(\frac{1}{c}(E+G)\right)= c \,
f\!\left(\frac{1}{c}E\right)+c\,f\!\left(\frac{1}{c}G\right)=
f(E)+f(G).
\ee

Let us now further extend $f$'s domain to the full space ${\cal
O}_d$. This can be done by noting that any operator $H$ can be
written as the difference $H=E-G$ of two positive semi-definite
operators.  Therefore define $f(H)\equiv f(E)-f(G)$, from which it
also follows that $f(-G)=-f(G)$. To see that this definition is
unique suppose there are four operators $E_1$, $E_2$, $G_1$, and
$G_2$, such that $H=E_1-G_1=E_2-G_2$.  It follows that
$E_1+G_2=E_2+G_1$. Applying $f$ (as extended in the previous
paragraph) to this equation, we obtain $f(E_1)+f(G_2)=f(E_2)+f(G_1)$
so that $f(E_1)-f(G_1)=f(E_2)-f(G_2)$. Finally, with this new
extension, full linearity can be checked immediately. This completes
the proof as far as the (complex) rational number field is
concerned:  Because $f$ extends uniquely to a linear functional on
${\cal O}_d$, we can indeed go through the steps of
Eqs.~(\ref{Hortense}) through (\ref{HollyHobie}) without worry.
\normalsize
\eq

There are two things that are significant about this much of the
proof.  First, in contrast to Gleason's original theorem, there is
nothing to bar the same logic from working when $d=2$.  This is quite
nice because much of the community has gotten in the habit of
thinking that there is nothing particularly ``quantum mechanical''
about a single qubit.  Indeed, because orthogonal projectors on
${\cal H}_2$ can be mapped onto antipodes of the Bloch sphere, it is
known that the measurement-outcome statistics for any standard
measurement can be mocked-up through a noncontextual hidden-variable
theory.  What this result shows is that that simply is not the case
when one considers the full set of POVMs as one's potential
measurements.

The other important thing is that the theorem works for Hilbert
spaces over the rational number field:  one does not need to invoke
the full power of the continuum. This contrasts with the surprising
result of Meyer~\cite{Meyer99} that the standard Gleason theorem
fails in such a setting.  The present theorem hints at a kind of
resiliency to the structure of quantum mechanics that falls through
the mesh of the standard Gleason result: The probability rule for
POVMs does not actually depend so much upon the detailed workings of
the number field.

The final step of the proof, indeed, is to show that nothing goes
awry when we go the extra step of reinstating the continuum.
\begin{quote}
In other words, we need to show that the function $f$ (now defined
on the set ${\cal E}_d$ complex operators) is a continuous function.
This comes about in simple way from $f$'s additivity. Suppose for
two positive semi-definite operators $E$ and $G$ that $E\le G$
(i.e., $G-E$ is positive semi-definite). Then trivially there exists
a positive semi-definite operator $H$ such that $E+H=G$ and through
which the additivity of $f$ gives $f(E)\le f(G)$.  Let $c$ be an
irrational number, and let $a_n$ be an increasing sequence and $b_n$
a decreasing sequence of rational numbers that both converge to $c$.
It follows for any positive semi-definite operator $E$, that
\be
f(a_n E)\le f(cE) \le f(b_n E)\;,
\ee
which implies
\be
a_n f(E)\le f(cE) \le b_n f(E)\;.
\ee

Since $\lim a_n f(E)$ and $\lim b_n f(E)$ are identical, by the
``pinching theorem'' of elementary calculus, they must equal
$f(cE)$. This establishes that we can consistently define
\be
f(cE)=cf(E)\;.
\ee
Reworking the extensions of $f$ in the last inset (but with this
enlarged notion of homogeneity), one completes the proof in a
straightforward manner.
\end{quote}

Of course we are not getting something from nothing.  The reason the
present derivation is so easy in contrast to the standard proof is
that {\it mathematically\/} the assumption of POVMs as the basic
notion of measurement is significantly stronger than the usual
assumption.  {\it Physically}, though, I would say it is just the
opposite.  Why add extra restrictions to the notion of measurement
when they only make the route from basic assumption to practical
usage more circuitous than it need be?

Still, no assumption should be left unanalyzed if it stands a chance
of bearing fruit.  Indeed, one can ask what is so very compelling
about the noncontextuality property (of probability assignments) that
both Gleason's original theorem and the present version make use of.
Given the picture of measurement as a kind of invasive intervention
into the world, one might expect the very opposite. One is left
wondering why measurement probabilities do not depend upon the whole
context of the measurement interaction.  Why is $p(b)$ not of the
form $f(b, \{E_c\})$?  Is there any precedent for the usual
assumption?

\section{Whither Bayes' Rule?}
\begin{flushright}
\parbox{3.1in}{\footnotesize
\bq
And so you see I have come to doubt\\ All that I once held as true\\
I stand alone without beliefs\\ The only truth I know is you.\smallskip\\
\hspace*{\fill} --- {\it Paul Simon}, Kathy's Song
\eq
}
\end{flushright}

Quantum states are states of knowledge, not states of nature.  That
statement is the cornerstone of this paper.  Thus, in searching to
make sense of the remainder of quantum mechanics, one strategy ought
to be to seek guidance~\cite{CavesFuchsSchack01} from the most
developed avenue of ``knowledge theory'' to date---Bayesian
probability theory~\cite{Kyburg80,JaynesPosthumous,Bernardo94}.
Indeed, the very aim of Bayesian theory is to develop reliable
methods of reasoning and making decisions in the light of incomplete
knowledge.  To what extent does that structure mesh with the
seemingly independent structure of quantum mechanics? To what extent
are there analogies; to what extent distinctions?

This section is about turning a distinction into an analogy.  The
core of the matter is the manner in which states of knowledge are
updated in the two theories.  At first sight, they appear to be
quite different in character.  To see this, let us first explore how
quantum mechanical states change when information is gathered.

In older accounts of quantum mechanics, one often encounters the
``collapse postulate'' as a basic statement of the theory.  One hears
things like, ``Axiom 5:  Upon the completion of a measurement of a
Hermitian observable $H$, the system is left in an eigenstate of
$H$.''  In quantum information, however, it has become clear that it
is useful to broaden the notion of measurement, and with it, the
analysis of how a state can change in the process.  The foremost
reason for this is that the collapse postulate is simply not true in
general:  Depending upon the exact nature of the measurement
interaction, there may be any of a large set of possibilities for
the final state of a system.

The broadest (consistent) notion of state change arose in the theory
of ``effects and operations''~\cite{Kraus83}. The statement is this.
Suppose one's initial state for a quantum system is a density
operator $\rho$, and a POVM $\{E_b\}$ is measured on that system.
Then, according to this formalism, the state after the measurement
can be {\it any\/} state $\rho_b$ of the form
\be
\rho_b = \frac{1}{\tr(\rho E_b)}\sum_i A_{bi}\rho A_{bi}^\dagger\;,
\label{StinkFart}
\ee
where
\be
\sum_i A_{bi}^\dagger A_{bi} = E_b\;.
\label{ColdWarsawDay}
\ee
Note the immense generality of this formula.  There is no constraint
on the number of indices $i$ in the $A_{bi}$ and these operators
need not even be Hermitian.

The {\it usual\/} justification for this kind of generality---just as
in the case of the commonplace justification for the POVM
formalism---comes about by imagining that the measurement arises in
an indirect fashion rather than as a direct attack. In other words,
the primary system is pictured to interact with an ancilla first,
and only then subjected to a ``real'' measurement on the ancilla
alone. The trick is that one posits a kind of projection postulate
on the primary system due to this process. This assumption has a
much safer feel than the raw projection postulate since, after the
interaction, no measurement on the ancilla should cause a physical
perturbation to the primary system.

More formally, we can start out by following Eqs.~(\ref{MushMush})
and (\ref{Aluminium}), but in place of Eq.~(\ref{RibTie}) we must
make an assumption on how the system's state changes.  For this one
invokes a kind of
``projection-postulate-at-a-distance.''\footnote{David Mermin has
also recently emphasized this point in Ref.~\cite{Mermin01}.} Namely,
one takes
\be
\rho_b=\frac{1}{P(b)}\;{\rm tr}_{\scriptscriptstyle\rm
A}\!\left((I\otimes\Pi_b) U(\rho_{\scriptscriptstyle\rm
S}\otimes\rho_{\scriptscriptstyle\rm A})
U^\dagger(I\otimes\Pi_b)\right)\;.
\label{BurpAtNight}
\ee
The reason for invoking the partial trace is to make sure that any
hint of a state change for the ancilla remains unaddressed.

To see how expression (\ref{BurpAtNight}) makes connection to
Eq.~(\ref{StinkFart}), denote the eigenvalues and eigenvectors of
$\rho_{\scriptscriptstyle\rm A}$ by $\lambda_\alpha$ and
$|a_\alpha\rangle$ respectively. Then $\rho_{\scriptscriptstyle\rm
S}\otimes\rho_{\scriptscriptstyle\rm A}$ can be written as
\be
\rho_{\scriptscriptstyle\rm S}\otimes\rho_{\scriptscriptstyle\rm
A}=\sum_\alpha
\sqrt{\lambda_\alpha}\,|a_\alpha\rangle\rho_{\scriptscriptstyle\rm
S} \langle a_\alpha|\sqrt{\lambda_\alpha}\;,
\ee
and, expanding Eq.~(\ref{BurpAtNight}), we have
\bea
\rho_b
&=&
\frac{1}{P(b)}\sum_\beta\langle a_\beta| (I\otimes\Pi_b) U^\dagger
(\rho_{\scriptscriptstyle\rm S}\otimes\rho_{\scriptscriptstyle\rm
A}) U (I\otimes\Pi_b)|a_\beta\rangle
\nonumber
\\
&=&
\frac{1}{P(b)}\sum_{\alpha\beta}\left(\sqrt{\lambda_\alpha}\,\langle
a_\beta| (I\otimes\Pi_b) U^\dagger
|a_\alpha\rangle\right)\rho_{\scriptscriptstyle\rm S}\left(\langle
a_\alpha| U(I\otimes\Pi_b)|a_\beta\rangle\sqrt{\lambda_\alpha}\,
\right)\;. \rule{0mm}{8mm}
\eea
A representation of the form in Eq.~(\ref{StinkFart}) can be made by
taking
\be
A_{b\alpha\beta}=\sqrt{\lambda_\alpha}\,\langle a_\alpha|
U(I\otimes\Pi_b)|a_\beta\rangle
\label{GoToBed}
\ee
and lumping the two indices $\alpha$ and $\beta$ into the single
index $i$. Indeed, one can easily check that
Eq.~(\ref{ColdWarsawDay}) holds.\footnote{As an aside, it should be
clear from the construction in Eq.~(\ref{GoToBed}) that there are
many equally good representations of $\rho_b$.  For a precise
statement of the latitude of this freedom, see
Ref.~\cite{Schumacher96}.} This completes what we had set out to
show. However, just as with the case of the POVM $\{E_b\}$, one can
always find a way to reverse engineer the derivation:  Given a set of
$A_{bi}$, one can always find a $U$, a $\rho_{\scriptscriptstyle\rm
A}$, and set of $\Pi_b$ such that Eq.~(\ref{BurpAtNight}) becomes
true.

Of course the old collapse postulate is contained within the extended
formalism as a special case:  There, one just takes both sets
$\{E_b\}$ and $\{A_{bi}=E_b\}$ to be sets of orthogonal projectors.
Let us take a moment to think about this special case in isolation.
What is distinctive about it is that it captures in the extreme a
common folklore associated with the measurement process. For it
tends to convey the image that measurement is a kind of gut wrenching
violence: In one moment the state is a
$\rho=|\psi\rangle\langle\psi|$, while in the very next it is a
$\Pi_i=|i\rangle\langle i|$. Moreover, such a wild transition need
depend upon no details of $|\psi\rangle$ and $|i\rangle$; in
particular the two states may even be almost orthogonal to each
other. In density-operator language, there is no sense in which
$\Pi_i$ is contained in $\rho$: the two states are simply in
distinct places of the operator space.

Contrast this with the description of information gathering that
arises in Bayesian probability theory.  There, an initial state of
knowledge is captured by a probability distribution $p(h)$ for some
hypothesis $H$.  The way gathering a piece of data $d$ is taken into
account in assigning one's new state of knowledge is through Bayes'
conditionalization rule.  That is to say, one expands $p(h)$ in terms
of the relevant joint probability distribution and picks off the
appropriate term:
\bea
p(h) &=& \sum_d p(h,d)
\nonumber
\\
&=& \sum_d p(d)p(h|d)
\label{Orecchiette}
\\
&& \phantom{\sum_d p(d)p(} \downarrow \nonumber\\
p(h) &&\stackrel{d}{\longrightarrow} \phantom{\sum_d} p(h|d)\;,
\label{Lasagna}
\eea
where $p(h|d)$ satisfies the tautology
\be
p(h|d)=\frac{p(h,d)}{p(d)}\;.
\label{Macaroni}
\ee
How gentle this looks in comparison to quantum collapse!  When one
gathers new information, one simply refines one's old knowledge in
the most literal of senses.  It is not as if the new state is
incommensurable with the old.  {\it It was always there}; it was just
initially averaged in with various other potential states of
knowledge.

Why does quantum collapse not look more like Bayes' rule?  Is
quantum collapse really a more violent kind of change, or might it be
an artifact of a problematic representation?  By this stage, it
should come as no surprise to the reader that dropping the ancilla
from our image of generalized measurements will be the first step to
progress.  Taking the transition from $\rho$ to $\rho_b$ in
Eqs.~(\ref{StinkFart}) and (\ref{ColdWarsawDay}) as the basic
statement of what quantum measurement {\it is\/} is a good starting
point.

To accentuate a similarity between Eq.~(\ref{StinkFart}) and Bayes'
rule, let us first contemplate cases of it where the index $i$ takes
on a single value.  Then, we can conveniently drop that index and
write
\be
\rho_b = \frac{1}{P(b)}A_b\rho A_b^\dagger\;,
\label{SlopFart}
\ee
where
\be
E_b = A_b^\dagger A_b\;.
\label{GrayJerseyDay}
\ee
In a loose way, one can say that measurements of this sort are the
most efficient they can be for a given POVM $\{E_b\}$:  For, a
measurement interaction with an explicit $i$-dependence may be viewed
as ``more truly'' a measurement of a finer-grained POVM that just
happens to throw away some of the information it gained.   Let us
make this point more precise.

Notice that Bayes' rule has the property that one's uncertainty
about a hypothesis can be expected to decrease upon the acquisition
of data.  This can be made rigorous, for instance, by gauging
uncertainty with the Shannon entropy function \cite{Cover91},
\be
S(H) = - \sum_h p(h)\log p(h)\;.
\ee
This number is bounded between 0 and the logarithm of the number of
hypotheses in $H$, and there are several reasons to think of it as a
good measure of uncertainty. Perhaps the most important of these is
that it quantifies the number of {\bf YES-NO} questions one can
expect to ask per instance of $H$, if one's only means to ascertain
the outcome is from a colleague who knows the actual
result~\cite{Ash65}. Under this quantification, the lower the Shannon
entropy, the more predictable a measurement's outcomes.

Because the function $f(x)=-x\log x$ is concave on the interval
$[0,1]$, it follows that,
\bea
S(H) &=& - \sum_h \left( \sum_d p(d)p(h|d)\right)\log \left( \sum_d
p(d)p(h|d)\right)
\nonumber
\\
&\ge&
-\sum_d p(d)\sum_h p(h|d)\log p(h|d)\;.
\nonumber
\\
&=&
\sum_d p(d) S(H|d)
\nonumber
\\
&\equiv&
S(H|D)\;.
\label{HungryBelly}
\eea

Indeed we hope to find a similar statement for how the result of
efficient quantum measurements decrease uncertainty or
impredictability. But, what can be meant by a decrease of uncertainty
through quantum measurement? I have argued strenuously that the
information gain in a measurement cannot be information about a
preexisting reality.  The way out of the impasse is simple: The
uncertainty that decreases in quantum measurement is the uncertainty
one expects for the results of potential future measurements.

There are at least two ways of quantifying this that are worthy of
note. The first has to do with the von Neumann entropy of a density
operator $\rho$:
\be
S(\rho) = - {\rm tr}\,\rho\log\rho
=-\sum_{k=1}^d\lambda_k\log\lambda_k\;,
\label{Lahti}
\ee
where the $\lambda_k$ signify the eigenvalues of $\rho$.  (We use
the convention that $\lambda\log\lambda=0$ whenever $\lambda=0$ so
that $S(\rho)$ is always well defined.)

The intuitive meaning of the von Neumann entropy can be found by
first thinking about the Shannon entropy. Consider any standard
measurement $\cal P$ consisting of $d$ one-dimensional orthogonal
projectors $\Pi_i$. The Shannon entropy for the outcomes of this
measurement is given by
\be
H({\cal P})=-\sum_{i=1}^d \big({\rm tr}\rho\Pi_i\big)\log\big({\rm
tr}\rho\Pi_i\big)\;.
\ee
A natural question to ask is:  With respect to a given density
operator $\rho$, which measurement $\cal P$ will give the most
predictability over its outcomes? As it turns out, the answer is any
$\cal P$ that forms a set of eigenprojectors for $\rho$
\cite{Wehrl78}. When this obtains, the Shannon entropy of the
measurement outcomes reduces to simply the von Neumann entropy of
the density operator. The von Neumann entropy, then, signifies the
amount of impredictability one achieves by way of a standard
measurement in a best case scenario.  Indeed, true to one's
intuition, one has the most knowledge by this account when $\rho$ is
a pure state---for then $S(\rho)=0$.  Alternatively, one has the
least knowledge when $\rho$ is proportional to the identity
operator---for then any measurement $\cal P$ will have outcomes that
are all equally likely.

The best case scenario for predictability, however, is a limited
case, and not very indicative of the density operator as a whole.
Since the density operator contains, in principle, all that can be
said about every possible measurement, it seems a shame to throw
away the vast part of that information in our considerations.

This leads to a second method for quantifying uncertainty in the
quantum setting. For this, we again rely on the Shannon information
as our basic notion of impredictability. The difference is we
evaluate it with respect to a ``typical'' measurement rather than
the best possible one. But typical with respect to what? The notion
of typical is only defined with respect to a given {\it measure\/} on
the set of measurements.

Regardless, there is a fairly canonical answer. There is a unique
measure $d\Omega_\Pi$ on the space of one-dimensional projectors
that is invariant with respect to all unitary operations.  That in
turn induces a canonical measure $d\Omega_{\cal P}$ on the space of
von Neumann measurements $\cal P$ \cite{Wootters90}. Using this
measure leads to the following quantity
\begin{eqnarray}
\overline{S}(\rho)&=&\int H(\Pi)\,d\Omega_{\cal P}
\nonumber
\\
&=& -d \int \big({\rm tr}\rho\Pi\big)\log\big({\rm
tr}\rho\Pi\big)\,d\Omega_\Pi\;,
\end{eqnarray}
which is intimately connected to the so-called quantum
``subentropy'' of Ref.~\cite{Jozsa94}.  This mean entropy can be
evaluated explicitly in terms of the eigenvalues of $\rho$ and takes
on the expression
\be
\overline{S}(\rho)=\frac{1}{\ln2}\left(\frac{1}{2}+
\frac{1}{3}+\cdots+\frac{1}{d}\right)+ Q(\rho)
\ee
where the subentropy $Q(\rho)$ is defined by
\be
Q(\rho)=-\sum_{k=1}^d\! \left(\prod_{i\ne
k}\frac{\lambda_k}{\lambda_k-\lambda_i}\right)\!\lambda_k\log\lambda_k\;.
\label{Mittelstaedt}
\ee
In the case where $\rho$ has degenerate eigenvalues,
$\lambda_l=\lambda_m$ for $l\ne m$, one need only reset them to
$\lambda_l+\epsilon$ and $\lambda_m-\epsilon$ and consider the limit
as $\epsilon\rightarrow0$.  The limit is convergent and hence
$Q(\rho)$ is finite for all $\rho$.  With this, one can also see
that for a pure state $\rho$, $Q(\rho)$ vanishes. Furthermore, since
$\overline{S}(\rho)$ is bounded above by $\log d$, we know that
\be
0\le Q(\rho)\le\log d - \frac{1}{\ln2}\!\left(\frac{1}{2}+
\cdots+\frac{1}{d}\right)\le\frac{1-\gamma}{\ln 2}\;,
\ee
where $\gamma$ is Euler's constant.  This means that for any $\rho$,
$Q(\rho)$ never exceeds approximately 0.60995 bits.

The interpretation of this result is the following.  Even when one
has maximal information about a quantum system---i.e., one has a
pure state for it---one can predict almost nothing about the outcome
of a typical measurement \cite{Caves96}. In the limit of large $d$,
the outcome entropy for a typical measurement is just a little over a
half bit away from its maximal value.  Having a mixed state for a
system, reduces one's predictability even further, but indeed not by
that much: The small deviation is captured by the function in
Eq.~(\ref{Mittelstaedt}), which becomes a quantification of
uncertainty in its own right.

The way to get at a quantum statement of Eq.~(\ref{HungryBelly}) is
to make use of the fact that $S(\rho)$ and $Q(\rho)$ are both
concave in the variable $\rho$ \cite{Fuchs00b}.  That is, for either
function, we have
\be
F\big(t\rho_0+(1-t)\rho_1\big)\ge t F(\rho_0) +(1-t)F(\rho_1)\;,
\label{BingoBopp}
\ee
for any density operators $\rho_0$ and $\rho_1$ and any real number
$t\in[0,1]$.  However, the result does not arise in the trivial
fashion it did for classical case of Eq.~(\ref{HungryBelly}).  This
is because generally, as already emphasized,
\be
\rho\ne \sum_b P(b) \rho_b
\ee
for $\rho_b$ defined as in Eq.~(\ref{SlopFart}).  One must be
slightly more roundabout.

The key is in noticing that
\bea
\rho &=& \rho^{1/2} I \rho^{1/2}
\nonumber
\\
&=& \sum_b \rho^{1/2}E_b \rho^{1/2}
\nonumber
\\
&=& \sum_b P(b) \tilde\rho_b
\eea
where
\be
\tilde\rho_b = \frac{1}{P(b)}\, \rho^{1/2}E_b \rho^{1/2} =
\frac{1}{P(b)}\, \rho^{1/2} A_b^\dagger A_b \rho^{1/2}\;.
\ee
What is special about this decomposition of $\rho$ is that for each
$b$, $\rho_b$ and $\tilde\rho_b$ have the same eigenvalues.  This
follows since $X^\dagger X$ and $X X^\dagger$ have the same
eigenvalues, for any operator $X$.  In the present case, setting
$X=A_b \rho^{1/2}$ does the trick.  Using the fact that both
$S(\rho)$ and $Q(\rho)$ depend only upon the eigenvalues of $\rho$
we obtain:
\bea
S(\rho) &\ge& \sum_b P(b) S(\rho_b)
\\
Q(\rho) &\ge& \sum_b P(b) Q(\rho_b)\;,
\eea
as we had been hoping for.  Thus, in performing an efficient quantum
measurement of a POVM $\{E_b\}$, an observer can expect to be left
with less uncertainty than he started with.\footnote{By differing
methods, a strengthening of this result in terms of a majorization
property can be found in Refs.~\cite{Fuchs00b} and
\cite{Nielsen00b}.}

In this sense, quantum ``collapse'' does indeed have some of the
flavor of Bayes' rule.  But we can expect more, and the derivation
above hints at just the right ingredient: $\rho_b$ and
$\tilde\rho_b$ have the same eigenvalues!  To see the impact of
this, let us once again explore the content of Eqs.~(\ref{SlopFart})
and (\ref{GrayJerseyDay}).  A common way to describe their meaning is
to use the operator polar-decomposition theorem \cite{Schatten60} to
rewrite Eq.~(\ref{SlopFart}) in the form
\be
\rho_b = \frac{1}{P(b)}\, U_b E_b^{1/2} \rho E_b^{1/2} U_b^\dagger\;,
\ee
where $U_b$ is a unitary operator. Since---subject only to the
constraint of efficiency---the operators $A_b$ are not determined any
further than Eq.~(\ref{GrayJerseyDay}), $U_b$ can be {\it any\/}
unitary operator whatsoever.  Thus, a customary way of thinking of
the state-change process is to break it up into two conceptual
pieces. First there is a ``raw collapse'':
\be
\rho\longrightarrow \sigma_b=\frac{1}{P(b)}\, E_b^{1/2} \rho
E_b^{1/2}\;.
\ee
Then, subject to the details of the measurement interaction and the
particular outcome $b$, one imagines the measuring device enforcing a
further kind of ``feedback'' on the measured system:
\be
\sigma_b\longrightarrow\rho_b=U_b\sigma_b U_b^\dagger\;.
\ee
But this break down of the transition is a purely conceptual game.

Since the $U_b$ are arbitrary to begin with, we might as well break
down the state-change process into the following (nonstandard)
conceptual components. First one imagines an observer refining his
initial state of knowledge and simply plucking out a term
corresponding to the ``data'' collected:
\bea
\rho &=& \sum_b P(b) \tilde\rho_b
\\
&& \phantom{\sum_b P(b)}\! \downarrow \nonumber
\\
\rho &&\stackrel{b}{\longrightarrow} \phantom{P(b)}\!\!\!
\tilde\rho_b\;.
\label{CoolEveningAir}
\eea
Finally, there may be a further ``mental readjustment'' of the
observer's knowledge, taking into account details both of the
measurement interaction and the observer's initial state of
knowledge. This is enacted via some (formal) unitary operation $V_b$:
\be
\tilde\rho_b \longrightarrow \rho_b = V_b \tilde\rho_b V_b^\dagger\;.
\ee
Putting the two processes together, one has the same result as the
usual picture.

The resemblance between the process in Eq.~(\ref{CoolEveningAir}) and
the classical Bayes' rule of Eq.~(\ref{Lasagna}) is
unmistakable.\footnote{Earlier allusions to a resemblance between
quantum collapse and Bayes' rule can be found in Ref.~\cite{Bub77}.}
By this way of viewing things, quantum collapse is indeed not such a
violent state of affairs after all.  Quantum measurement is nothing
more, and nothing less, than a refinement and a readjustment of
one's state of knowledge.  More general state changes of the form
Eq.~(\ref{StinkFart}) come about similarly, but with a further step
of coarse-graining (i.e., throwing away information that was in
principle accessible).

Let us look at two limiting cases of efficient measurements.  In the
first, we imagine an observer whose initial state of knowledge
$\rho=|\psi\rangle\langle\psi|$ is a maximal state of knowledge.  By
this account, no measurement whatsoever can {\it refine\/} that
state of knowledge.  This follows because, no matter what $\{E_b\}$
is,
\be
\rho^{1/2}E_b \rho^{1/2} = P(b)|\psi\rangle\langle\psi|\;.
\ee
The only state change that can come about from such a measurement
must be purely of the ``mental readjustment'' sort:  We learn nothing
new; we just change what we can predict as a consequence of
experimental intervention.  In particular, when the POVM is an
orthogonal set of projectors $\{\Pi_i=|i\rangle\langle i|\}$ and the
state-change mechanism is the von Neumann collapse postulate, this
simply corresponds to a readjustment according to the unitary
operators
\be
U_i=|i\rangle\langle\psi|\;.
\ee

At the opposite end of things, we can contemplate measurements that
have no causal connection at all to the system being measured.  This
could come about, for instance, by interacting with one side of an
entangled pair of systems and using the consequence of that
intervention to update one's knowledge about the other side.  In
such a case, one can show that the state change is purely of the
refinement variety (with no further mental readjustment).  For
instance, consider a pure state $|\psi^{\scriptscriptstyle
AB}\rangle$ whose Schmidt decomposition takes the form
\be
|\psi^{\scriptscriptstyle AB}\rangle = \sum_i \sqrt{\lambda_i}
|a_i\rangle|b_i\rangle\;.
\ee
An efficient measurement on the $A$ side of this leads to a state
update of the form
\be
|\psi^{\scriptscriptstyle AB}\rangle\langle \psi^{\scriptscriptstyle
AB}|\, \longrightarrow\, (A_b\otimes I) |\psi^{\scriptscriptstyle
AB}\rangle\langle \psi^{\scriptscriptstyle AB}| (A_b^\dagger\otimes
I)\;.
\ee
Tracing out the $A$ side, then gives
\bea
\!\!\!\!\!\!\!\!\!\!\!\!\!\! \tr_{\scriptscriptstyle\rm A}
\Big(A_b\otimes I |\psi^{\scriptscriptstyle AB}\rangle\langle
\psi^{\scriptscriptstyle AB}| A_b^\dagger\otimes I\Big)
&=&
\sum_{ijk}\sqrt{\lambda_j}\sqrt{\lambda_k}\langle a_i|A_b\otimes I
|a_j\rangle|b_j\rangle\langle a_k|\langle b_k|A_b^\dagger\otimes
I|a_i\rangle
\nonumber
\\
&=&
\sum_{ijk}\sqrt{\lambda_j}\sqrt{\lambda_k}\langle a_k|A_b^\dagger
|a_i\rangle\langle a_i|A_b|a_j\rangle |b_j\rangle\langle b_k|
\nonumber
\\
&=&
\sum_{jk}\sqrt{\lambda_j}\sqrt{\lambda_k}\langle a_k|A_b^\dagger
A_b|a_j\rangle |b_j\rangle\langle b_k|
\nonumber
\\
&=&
\sum_{jk}\sqrt{\lambda_j}\sqrt{\lambda_k}\langle b_k|UA_b^\dagger
A_bU^\dagger|b_j\rangle |b_j\rangle\langle b_k|
\nonumber
\\
&=&
\sum_{jk}\sqrt{\lambda_j}\sqrt{\lambda_k}\langle
b_j|\Big(UA_b^\dagger A_bU^\dagger\Big)^{\scriptscriptstyle\rm
T}|b_k\rangle |b_j\rangle\langle b_k|
\nonumber
\\
&=&
\rho^{1/2} \Big(UA_b^\dagger A_bU^\dagger\Big)^{\scriptscriptstyle\rm
T} \rho^{1/2}
\eea
where $\rho$ is the initial quantum state on the $B$ side, $U$ is
the unitary operator connecting the $|a_i\rangle$ basis to the
$|b_i\rangle$ basis, and $^{\scriptscriptstyle\rm T}$ represents
taking a transpose with respect to the $|b_i\rangle$ basis.  Since
the operators
\be
F_b=\Big(UA_b^\dagger A_bU^\dagger\Big)^{\scriptscriptstyle\rm T}
\ee
go together to form a POVM, we indeed have the claimed result.

In summary, the lesson here is that it turns out to be rather easy to
think of quantum collapse as a noncommutative variant of Bayes'
rule.  In fact it is just in this that one starts to get a feel for
(at least) a partial reason for Gleason's noncontextuality
assumption.  In the setting of classical Bayes' conditionalization
we have just that:  The probability of the transition
$p(h)\longrightarrow p(h|d)$ is governed solely by the local
probability $p(d)$. The transition does not care about how we have
partitioned the rest of the potential transitions.  That is, it does
not care whether $d$ is embedded in a two outcome set $\{d,\neg d\}$
or whether it is embedded in a three outcome set, $\{d,e,\neg (d\vee
e)\}$, etc. Similarly with the quantum case. The probability for a
transition from $\rho$ to $\rho_0$ cares not whether our refinement
is of the form
\be
\rho = P(0)\rho_0 + \sum_{b=1}^{17} P(b)\rho_b\quad \qquad \mbox{or
of the form}\quad \qquad \rho = P(0)\rho_0 + P(18)\rho_{18}\;,
\ee
as long as
\be
P(18)\rho_{18} = \sum_{b=1}^{17} P(b)\rho_b
\ee
What could be a simpler generalization of Bayes' rule?

Indeed, leaning on that, we can restate the discussion of the
``measurement problem'' at the beginning of Section 4 in slightly
more technical terms.  Go back to the classical setting of
Eqs.~(\ref{Orecchiette}) and (\ref{Macaroni}) where an agent has a
probability distribution $p(h,d)$ over two sets of hypotheses.
Marginalizing over the possibilities for $d$, one obtains the agent's
initial state of knowledge $p(h)$ for the hypothesis $h$. If he
gathers an explicit piece of data $d$, he should use Bayes' rule to
update his knowledge about $h$ to $p(h|d)$.

The question is this: Is the transition
\be
p(h) \longrightarrow p(h|d)
\ee
a mystery we should contend with?  If someone asked for a {\it
physical description\/} of that transition, would we be able to give
an explanation? After all, one value for $h$ is true and always
remains true: there is no transition in it. One value for $d$ is
true and always remains true:  there is no transition in it. The only
transition is in the {\it knowledge\/} $p(h)$.  To put the issue into
perspective for the quantum measurement problem, let us ask: Should
we not have a detailed theory of how the brain works before we can
trust in the validity of Bayes' rule?

The answer is, ``Of course not!''  Bayes' rule, and with it all of
probability theory, is an intellectual construct that stands beyond
the details of physics. George Boole called probability theory a
{\it law of thought\/}~\cite{Boole58}.  Its calculus specifies the
optimal way an agent should reason and make decisions when faced with
incomplete information.  In this way, probability theory is but a
generalization of Aristotelian logic\footnote{In addition to
Ref.~\cite{JaynesPosthumous}, many further materials concerning this
point of view can be downloaded from the {\sl Probability Theory As
Extended Logic\/} web site maintained by G.~L. Bretthorst, {\tt
http://bayes. wustl.edu/}.}---a construct very few would accept as
being tied to the explicit details of the physical world.\footnote{We
have, after all, used simple Aristotelian logic in making deductions
from {\it all\/} our physical theories to date: from Aristotle's
physics to quantum mechanics to general relativity and even string
theory.}

The formal similarities between Bayes' rule and quantum collapse may
be telling us how to finally cut the Gordian knot of the measurement
problem. Namely, it may be telling us that it is simply not a problem
at all! Indeed, drawing on the analogies between the two theories,
one is left with a spark of insight:  perhaps the better part of
quantum mechanics is simply ``law of thought''~\cite{Fuchs01}.
Perhaps the structure of the theory denotes the optimal way to
reason and make decisions in light of {\it some\/} fundamental
situation, waiting to be ferreted out in a more satisfactory fashion.

This much we know:  That ``fundamental situation''---whatever it
is---must be an ingredient Bayesian probability theory does not
have. There must be something to drive a wedge between the two
theories. Probability theory alone is too general of a structure.
Narrowing it will require input from the world about us.

\section{Wither Entanglement?}
\begin{flushright}
\parbox{4.0in}{\footnotesize
\bq
When two systems, of which we know the states by their respective
representatives, enter into temporary physical interaction due to
known forces between them, and when after a time of mutual influence
the systems separate again, then they can no longer be described in
the same way as before, viz.\ by endowing each of them with a
representative of its own.  I would not call that {\it one\/} but
rather {\it the\/} characteristic trait of quantum mechanics, the one
that enforces its entire departure from classical lines of thought.
By the interaction the two representatives (or $\psi$-functions) have
become entangled.\\
\hspace*{\fill} --- {\it Erwin Schr\"odinger}, 1935
\eq
}
\end{flushright}

Quantum entanglement certainly gets a load of airplay these days.  By
most accounts it is the main ingredient in quantum information
theory and quantum computing~\cite{Jozsa98}, and it is the main
mystery of the quantum foundations~\cite{Mermin90}. But what is it?
Where does it come from?

The predominant purpose it has served in this paper has been as a
kind of background.  For it, more than any other ingredient in
quantum mechanics, has clinched the issue of ``information about
what?''~in the author's mind:  That information cannot be about a
preexisting reality (a hidden variable) unless we are willing to
renege on our very reason for rejecting the quantum state's
objective reality in the first place. What I am alluding to here is
the conjunction of the Einstein argument reported in Section 3 and
the phenomena of the Bell inequality violations by quantum
mechanics. Putting those points together gave us that the
information symbolized by a $|\psi\rangle$ must be information about
the potential consequences of our interventions into the world.

But, now I would like to turn the tables and ask whether the
structure of our potential interventions---the POVMs---can tell us
something about the origin of entanglement.  Could it be that the
concept of entanglement is just a minor addition to the much deeper
point that measurements correspond to refinements of density
operators (i.e., the substance of the two preceding sections)?

The technical translation of this question is, why do we combine
systems according to the tensor product rule?  There are certainly
innumerable ways to combine two Hilbert spaces ${\cal H}_A$ and
${\cal H}_B$ to obtain a third ${\cal H}_{AB}$.  We could take the
direct sum of the two spaces ${\cal H}_{AB}={\cal H}_{A}\oplus{\cal
H}_{B}$. We could take their Grassmann product ${\cal H}_{AB}={\cal
H}_{A}\wedge{\cal H}_{B}$~\cite{Bhatia97}.  We could take scads of
other things. But instead we take their tensor product,
\be
{\cal H}_{AB}={\cal H}_{A}\otimes{\cal H}_{B}\;.
\ee
Why?

Could it arise from the selfsame considerations we have already made
our mainstay---from a noncontextuality property for
measurement-outcome probabilities? The answer is yes, and the
theorem I am about demonstrate owes much in inspiration to
Ref.~\cite{Wallach00}.

Here is the scenario. Suppose we have two quantum systems, and we can
make a measurement on each.\footnote{This, one might think, is the
very essence of having {\it two\/} systems rather than one---that we
can probe them independently.} On the first, we can measure any POVM
on the $d_{\rm A}$-dimensional Hilbert space ${\cal H}_A$; on the
second, we can measure any POVM on the $d_{\rm B}$-dimensional
Hilbert space ${\cal H}_B$. Moreover, suppose we may condition the
second measurement on the nature and the outcome of the first, and
vice versa. That is to say---walking from $A$ to $B$---we could
first measure $\{E_i\}$ on $A$, and then, depending on the outcome
$i$, measure $\{F^i_j\}$ on $B$. Similarly---walking from $B$ to
$A$---we could first measure $\{F_j\}$ on $B$, and then, depending
on the outcome $j$, measure $\{E^j_i\}$ on $A$. So that we have valid
POVMs, we must have
\be
\sum_i E_i = I \qquad\mbox{and}\qquad \sum_j F^i_j = I \quad\forall\,
i\;,
\label{Herme}
\ee
and
\be
\sum_i E^j_i = I \quad\forall\, j \qquad\mbox{and}\qquad \sum_j F_j =
I\;,
\label{Neutics}
\ee
for these sets of operators.  Let us denote by $S_{ij}$ an {\it
ordered pair\/} of operators, either of the form $(E_i, F^i_j)$ or
of the form $(E^j_i, F_j)$, as appearing above.  Let us call a set
of such operators $\{S_{ij}\}$ a {\it locally-measurable POVM tree}.

Suppose now that---just as with the POVM-version of Gleason's theorem
in Section 4---the joint probability $P(i,j)$ for the outcomes of
such a measurement should not depend upon which tree $S_{ij}$ is
embedded in:  This is essentially the same assumption we made there,
but now applied to local measurements on the separate systems. In
other words, let us suppose there exists a function
\be
f:{\cal E}_{d_{\scriptscriptstyle\rm A}}\times {\cal
E}_{d_{\scriptscriptstyle\rm B}} \longrightarrow [0,1]
\label{Ersatz}
\ee
such that
\be
\sum_{ij} f(S_{ij}) = 1
\label{HamAndEggs}
\ee
whenever the $S_{ij}\/$ satisfy either Eq.~(\ref{Herme}) or
Eq.~(\ref{Neutics}).

Note in particular that Eq.~(\ref{Ersatz}) makes no use of the
tensor product:  The domain of $f$ is the {\it Cartesian product\/}
of the two sets ${\cal E}_{d_{\scriptscriptstyle\rm A}}$ and ${\cal
E}_{d_{\scriptscriptstyle\rm B}}$.  The notion of a {\it local\/}
measurement on the separate systems is enforced by the requirement
that the ordered pairs $S_{ij}$ satisfy the side conditions of
Eqs.~(\ref{Herme}) and (\ref{Neutics}).  This, of course, is not the
most general kind of local measurement one can imagine---more
sophisticated measurements could involve multiple ping-pongings
between $A$ and $B$ as in Ref.~\cite{Bennett99}---but the present
restricted class is already sufficient for fixing the probability
rule for local measurements.

The {\it theorem\/} is this: If $f$ satisfies Eqs.~(\ref{Ersatz}) and
(\ref{HamAndEggs}) for all locally-measurable POVM trees, then there
exists a density operator $\tilde\rho$ on ${\cal H}_{A}\otimes{\cal
H}_{B}$ such that
\be
f(E,F)=\tr\Big(\tilde{\rho}(E\otimes F)\Big)\;.
\ee
If ${\cal H}_{A}$ and ${\cal H}_{B}$ are defined over the field of
complex numbers, then $\tilde\rho$ is unique.  Uniqueness does not
hold, however, if the underlying field is the real numbers.

The proof of this statement is almost a trivial extension of the
proof in Section 4.  One again starts by showing additivity, but
this time in the two variables $E$ and $F$ separately.  For
instance, for a fixed $E\in{\cal E}_{d_{\scriptscriptstyle\rm A}}$,
define
\be
g_E(F) = f(E,F)\;,
\ee
and consider two locally-measurable POVM trees
\be
\{(I-E,F_i), (E, G_\alpha)\} \qquad\mbox{and}\qquad \{(I-E,F_i), (E,
H_\beta)\}\;,
\ee
where $\{F_i\}$, $\{G_\alpha\}$, and $\{H_\beta\}$ are arbitrary
POVMs on ${\cal H}_B$.  Then Eq.~(\ref{HamAndEggs}) requires that
\be
\sum_i g_{I\mbox{-}E}(F_i)+\sum_\alpha g_E(G_\alpha) = 1
\ee
and
\be
\sum_i g_{I\mbox{-}E}(F_i)+\sum_\beta g_E(H_\beta) = 1\;.
\ee
From this it follows that,
\be
\sum_\alpha g_E(G_\alpha) = \sum_\beta g_E(H_\beta) = \mbox{const}.
\ee
That is to say, $g_E(F)$ is a frame function in the sense of Section
4.  Consequently, we know that we can use the same methods as there
to uniquely extend $g_E(F)$ to a linear functional on the complete
set of Hermitian operators on ${\cal H}_B$. Similarly, for fixed
$F\in{\cal E}_{d_{\scriptscriptstyle\rm B}}$, we can define
\be
h_F(E) = f(E,F)\;,
\ee
and prove that this function too can be uniquely extended to a linear
functional on the Hermitian operators on ${\cal H}_A$.

The linear extensions of $g_E(F)$ and $h_F(E)$ can be put together
in a simple way to give a full bilinear extension to the function
$f(E,F)$.  Namely, for any two Hermitian operators $A$ and $B$ on
${\cal H}_A$ and ${\cal H}_B$, respectively, let $A=\alpha_1 E_1 -
\alpha_2 E_2$ and $B=\beta_1 F_1 - \beta_2 F_2$ be decompositions
such that $\alpha_1,\alpha_2,\beta_1,\beta_2\ge0$, $E_1,E_2\in {\cal
E}_{d_{\scriptscriptstyle\rm A}}$, and $F_1,F_2\in {\cal
E}_{d_{\scriptscriptstyle\rm B}}$.  Then define
\be
f(A,B)\equiv \alpha_1 g_{E_1}(B) - \alpha_2 g_{E_2}(B)\;.
\ee
To see that this definition is unique, take any other decomposition
$A=\tilde\alpha_1 \tilde E_1 - \tilde\alpha_2 \tilde E_2$.  Then we
have
\bea
f(A,B) &=& \tilde \alpha_1 g_{\tilde E_1}(B) - \tilde \alpha_2
g_{\tilde E_2}(B)
\nonumber
\\
&=&
\tilde \alpha_1 f(\tilde E_1,B) - \tilde \alpha_2 f(\tilde E_2,B)
\nonumber
\\
&=&
\beta_1 \Big(\tilde \alpha_1 f(\tilde E_1,F_1)- \tilde \alpha_2 f(\tilde E_1,F_1)\Big) -
\beta_2 \Big(\tilde \alpha_1 f(\tilde E_1,F_2)- \tilde \alpha_2 f(\tilde E_2,F_2)\Big)
\nonumber
\\
&=&
\beta_1 h_{F_1}(A) - \beta_2 h_{F_2}(A)
\nonumber
\\
&=&
\beta_1 \Big(\alpha_1 f(E_1,F_1)- \alpha_2 f(E_1,F_1)\Big) -
\beta_2 \Big(\alpha_1 f(E_1,F_2)- \alpha_2 f(E_2,F_2)\Big)
\nonumber
\\
&=&
\alpha_1 f(E_1,B) - \alpha_2 f(E_2,B)
\nonumber
\\
&=&
\alpha_1 g_{E_1}(B) - \alpha_2 g_{E_2}(B)\;,
\eea
which is as desired.

With bilinearity for the function $f$ established, we have
essentially the full story~\cite{Bhatia97,MacLane67}.  For, let
$\{E_i\}$, $i=1,\ldots,d_{\rm A}^2$, be a complete basis for the
Hermitian operators on ${\cal H}_A$ and let $\{F_j\}$,
$j=1,\ldots,d_{\rm B}^2$, be a complete basis for the Hermitian
operators on ${\cal H}_B$.  If $E=\sum_i\alpha_i E_i$ and
$F=\sum_j\beta_j F_j$, then
\be
f(E,F)=\sum_{ij}\alpha_i\beta_j f(E_i,F_j)\;.
\ee
Define $\tilde\rho$ to be a linear operator on ${\cal
H}_{A}\otimes{\cal H}_{B}$ satisfying the $(d_{\rm A}d_{\rm B})^2$
linear equations
\be
\tr\Big(\tilde\rho(E_i\otimes F_j)\Big)=f(E_i,F_j)\;.
\label{BigGrump}
\ee
Such an operator always exists.  Consequently we have,
\bea
f(E,F) &=& \sum_{ij}\alpha_i\beta_j \tr\Big(\tilde\rho(E_i\otimes
F_j)\Big)
\nonumber
\\
&=&
\tr\Big(\tilde{\rho}(E\otimes F)\Big)\;.
\eea
Enforcing positivity and normalization for the function $f$ proves
the main point of the theorem.

For complex Hilbert spaces ${\cal H}_A$ and ${\cal H}_B$, the
uniqueness of $\tilde\rho$ comes about because the set $\{E_i\otimes
F_j\}$ forms a complete basis for the Hermitian operators on ${\cal
H}_{A}\otimes{\cal H}_{B}$.  For real Hilbert spaces, however, the
analog of the Hermitian operators are the symmetric operators.  The
dimensionality of the space of symmetric operators on a real Hilbert
space ${\cal H}_d$ is $\frac{1}{2}d(d+1)$, rather than the $d^2$ it
is for the complex case.  This means that in the steps above only
\be
\frac{1}{4} d_{\rm A}d_{\rm B}(d_{\rm A}+1)(d_{\rm B}+1)
\label{LoudTalk}
\ee
equations will appear in Eq.~(\ref{BigGrump}), whereas
\be
\frac{1}{2} d_{\rm A}d_{\rm B} (d_{\rm A}d_{\rm B}+1)
\label{Ulcer}
\ee
are needed to uniquely specify a $\tilde\rho$.  For instance take
$d_{\rm A}=d_{\rm B}=2$.  Then Eq.~(\ref{LoudTalk}) gives nine
equations, while Eq.~(\ref{Ulcer}) requires ten.

Let us emphasize the striking feature of this way of deriving the
tensor product rule for combining separate quantum systems:  It is
built on the very concept of local measurement.  There is nothing
``spooky'' or ``nonlocal'' about it; there is nothing in it
resembling ``passion at a distance''~\cite{ShimonyFest}.  Indeed,
one need not even consider probability assignments for the outcomes
of measurements of the ``nonlocality without entanglement''
variety~\cite{Bennett99} in order to uniquely fix the probability
rule. That is---to give an example on ${\cal H}_3\otimes{\cal
H}_3$---one need not consider standard measurements like
$\{E_b=|\psi_b\rangle\langle\psi_b|\}$, where
\bea
|\psi_1\rangle \!\! &=& \!\! |1\rangle |1\rangle \nonumber\\
|\psi_2\rangle = |0\rangle |0+1\rangle \quad\qquad && \,\quad\qquad
|\psi_6\rangle = |1+2\rangle |0\rangle \nonumber\\
|\psi_3\rangle = |0\rangle |0-1\rangle \quad\qquad && \,\quad \qquad
|\psi_7\rangle = |1-2\rangle |0\rangle\\
|\psi_4\rangle = |2\rangle |1+2\rangle \quad\qquad && \,\quad \qquad
|\psi_8\rangle = |0+1\rangle |2\rangle \nonumber\\
|\psi_5\rangle = |2\rangle |1-2\rangle \quad\qquad && \,\quad \qquad
|\psi_9\rangle = |0-1\rangle |2\rangle \nonumber
\eea
with $|0\rangle$, $|1\rangle$, and $|2\rangle$ forming an
orthonormal basis on ${\cal H}_3$, and
$|0+1\rangle=\frac{1}{\sqrt{2}}(|0\rangle+|1\rangle)$, etc.  This is
a measurement that takes neither the form of Eq.~(\ref{Herme}) nor
(\ref{Neutics}).  It stands out instead in that, even though all its
POVM elements are tensor product operators---i.e., they have no
quantum entanglement---it still {\it cannot\/} be measured by local
means, even with the elaborate ping-ponging strategies mentioned
earlier.

Thus, the tensor product rule, and with it quantum entanglement,
seems to be more a statement of locality than anything else.  It,
like the probability rule, is more a product of the structure of the
observables---that they are POVMs---combined with noncontextuality.
In searching for the secret ingredient to drive a wedge between
general Bayesian probability theory and quantum mechanics, it seems
that the direction {\it not\/} to look is toward quantum
entanglement. Perhaps the trick instead is to dig deeper into the
Bayesian toolbox.

\section{Unknown Quantum States?}
\begin{flushright}
\parbox{4.0in}{\footnotesize
\bq
My thesis, paradoxically, and a little pro\-vocatively, but nonetheless
genuinely, is simply this:
\begin{center}
QUANTUM STATES DO NOT EXIST.
\end{center}
The abandonment of superstitious beliefs about the existence of
Phlogiston, the Cosmic Ether, Absolute Space and Time, ..., or Fairies
and Witches, was an essential step along the road to scientific
thinking. The quantum state, too, if regarded as something endowed
with some kind of objective existence, is no less a misleading
conception, an illusory attempt to exteriorize or
materialize the information we possess.
\\
\hspace*{\fill} --- {\it the ghost of Bruno de Finetti}
\eq
}
\end{flushright}
\vspace{-.2in}

The hint of a more fruitful direction can be found by trying to make
sense of one of the most commonly used phrases in quantum information
theory. It is the {\it unknown quantum state}. There is hardly a
paper in quantum information that does not make use of it. Unknown
quantum states are teleported~\cite{Bennett93}, protected with
quantum error correcting codes~\cite{Shor95}, and used to check for
quantum eavesdropping~\cite{Bennett84}.  The list of uses grows each
day. But what can the term possibly mean?  In an information-based
interpretation of quantum mechanics, it is an overt oxymoron: If
quantum states, by their very definition, are states of knowledge
and not states of nature, then the state is {\it known\/} by
someone---at the very least, by the person who wrote it down.

Thus, if a phenomenon ostensibly invokes the concept of an unknown
state in its formulation, that unknown state had better be shorthand
for a more basic situation (even if that basic situation still awaits
a complete analysis).  This means that for any phenomenon using the
idea of an unknown quantum state in its description, we should
demand that either
\begin{enumerate}
\item
The owner of the unknown state---a further decision-making agent or
observer---be explicitly identified.  (In this case, the unknown
state is merely a stand-in for the unknown {\it state of
knowledge\/} of an essential player who went unrecognized in the
original formulation.)  Or,
\item
If there is clearly no further agent or observer on the scene, then
a way must be found to reexpress the phenomenon with the term
``unknown state'' completely banished from its formulation. (In this
case, the end-product of the effort will be a single quantum state
used for describing the phenomenon---namely, the state that actually
captures the describer's state of knowledge throughout.)
\end{enumerate}

\begin{figure} 
\begin{center}
\epsfxsize=8cm \epsfbox{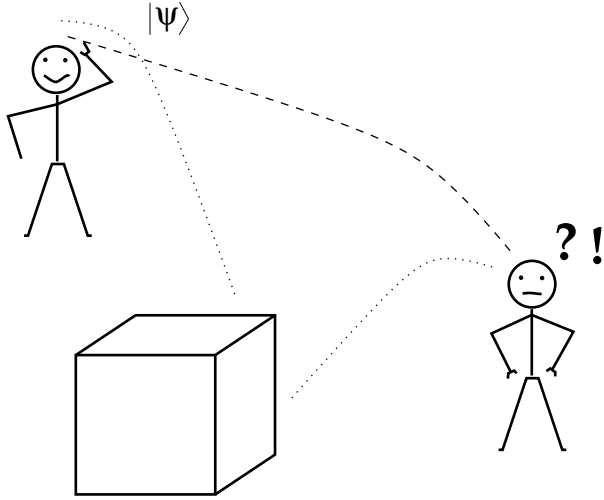} \bigskip\caption{\small What can
the term ``unknown state'' mean if quantum states are taken
exclusively to be states of knowledge rather than states of nature?
When we say that a system has an unknown state, must we always
imagine a further observer whose state of knowledge is symbolized by
some $|\psi\rangle$, and it is the identity of the symbol that we are
ignorant of?}
\end{center}
\end{figure}

This Section reports the work of Ref.~\cite{Caves01} and
\cite{Schack00}, where such a project is carried out for the
experimental practice of {\it quantum-state
tomography\/}~\cite{Vogel89}. The usual description of tomography is
this.  A device of some sort, say a nonlinear optical medium driven
by a laser, repeatedly prepares many instances of a quantum system,
say many temporally distinct modes of the electromagnetic field, in
a fixed quantum state $\rho$, pure or mixed~\cite{vanEnk01}.  An
experimentalist who wishes to characterize the operation of the
device or to calibrate it for future use may be able to perform
measurements on the systems it prepares even if he cannot get at the
device itself. This can be useful if the experimenter has some prior
knowledge of the device's operation that can be translated into a
probability distribution over states. Then learning about the state
will also be learning about the device. Most importantly, though,
this description of tomography assumes that the precise state $\rho$
is unknown.  The goal of the experimenter is to perform enough
measurements, and enough kinds of measurements (on a large enough
sample), to estimate the identity of $\rho$.

This is clearly an example where there is no further player on whom
to pin the unknown state as a state of knowledge.  Any attempt to
find such a missing player would be entirely artificial: Where would
the player be placed?  On the inside of the device the tomographer is
trying to characterize?\footnote{Placing the player here would be
about as respectable as George Berkeley's famous patch to his
philosophical system of idealism. The difficulty is captured
engagingly by a limerick of Ronald Knox and its anonymous reply:
\bq
There was a young man who said, ``God {\bf :} Must think it
exceedingly odd {\bf :} If he finds that this tree {\bf :} Continues
to be {\bf :} When there's no one about in the Quad.'' REPLY: ``Dear
Sir: Your astonishment's odd. {\bf :} I am always about in the Quad.
{\bf :} And that's why the tree {\bf :} Will continue to be, {\bf :}
Since observed by Yours faithfully, God.''\eq} The only available
course is the second strategy above---to banish the idea of the
unknown state from the formulation of tomography.

\begin{figure} \leavevmode
\begin{center}
\epsfxsize=9cm \epsfbox{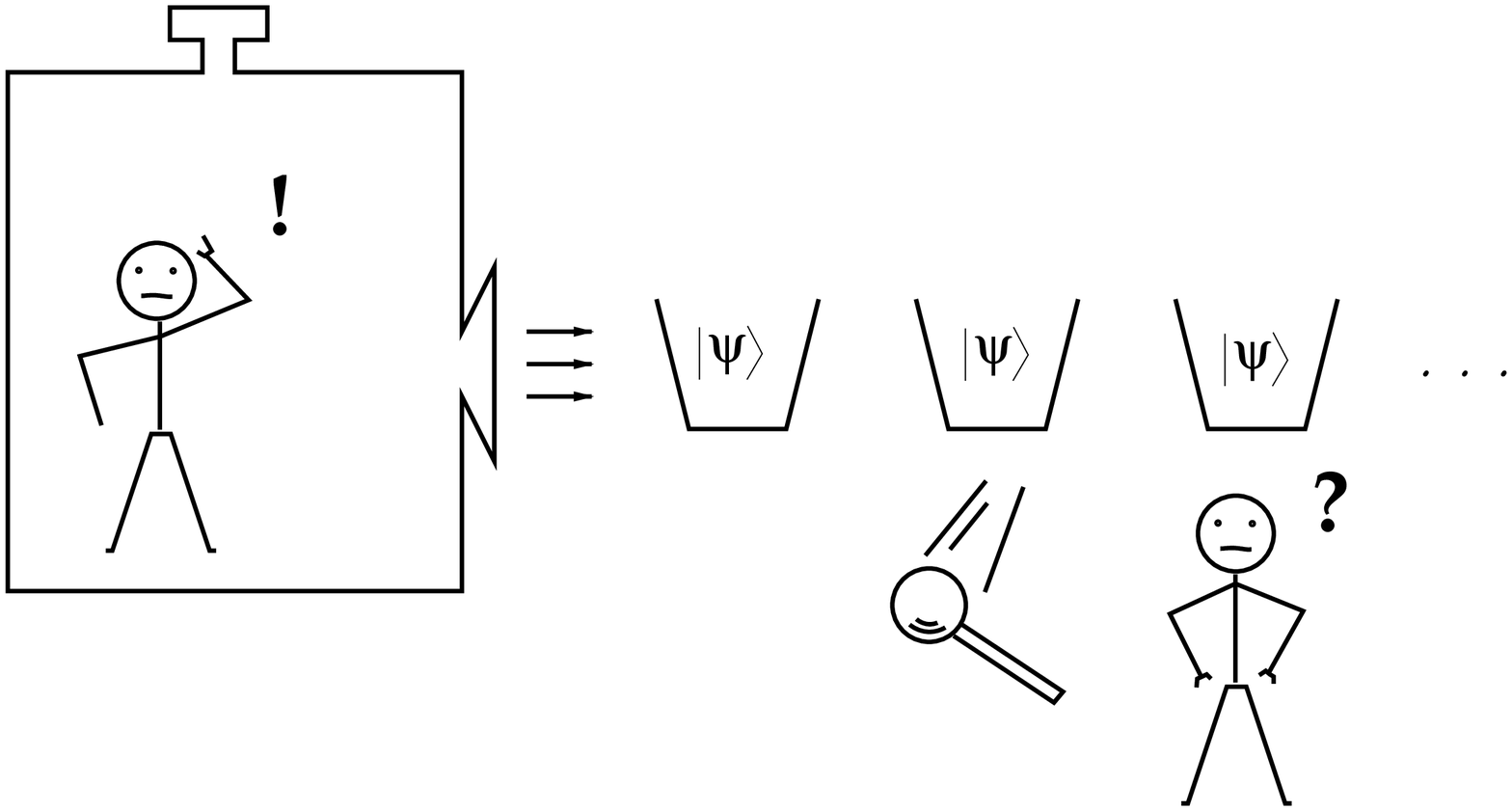} \bigskip\caption{\small To make
sense of quantum tomography, must we go to the extreme of imagining
a ``man in the box'' who has a better description of the systems
than we do?  How contrived our usage would be if that were so!}
\end{center}
\end{figure}

To do this, we once again take our cue from Bayesian probability
theory~\cite{Kyburg80,JaynesPosthumous,Bernardo94}. As emphasized
previously, in Bayesian theory probabilities---just like quantum
states---are not objective states of nature, but rather measures of
credible belief, reflecting one's state of knowledge. In light of
this, it comes as no surprise that one of the most overarching
Bayesian themes is to identify the conditions under which a set of
decision-making agents can come to a common belief or probability
assignment for a random variable even though their initial beliefs
may differ~\cite{Bernardo94}. Following that theme is the key to
understanding the essence of quantum-state tomography.

Indeed, classical Bayesian theory encounters almost precisely the
same problem as our unknown quantum state through the widespread use
of the phrase ``unknown probability'' in its domain.  This is an
oxymoron every bit as egregious as unknown state.

The procedure analogous to quantum-state tomography in Bayesian
theory is the estimation of an unknown probability from the results
of repeated trials on ``identically prepared systems.'' The way to
eliminate unknown probabilities from this situation was introduced
by Bruno de Finetti in the early 1930s \cite{DeFinetti1990}.  His
method was simply to focus on the equivalence of the repeated
trials---namely, that what is really of importance is that the
systems are indistinguishable as far as probabilistic predictions
are concerned.  Because of this, any probability assignment
$p(x_1,x_2,\ldots,x_N)$ for multiple trials should be symmetric
under permutation of the systems.  As innocent as this conceptual
shift may sound, de Finetti was able to use it to powerful effect.
For, with his {\it representation theorem}, he showed that any
multi-trial probability assignment that is permutation-symmetric for
an arbitrarily large number of trials---de Finetti called such
multi-trial probabilities {\it exchangeable\/}---is equivalent to a
probability for the ``unknown probabilities.''

Let us outline this in a little more detail. In an objectivist
description of $N$ ``identically prepared systems,'' the individual
trials are described by discrete random variables
$x_n\in\{1,2,\ldots,k\}$, $n=1,\ldots,N$, and the probability in the
multi-trial hypothesis space is given by an independent identically
distributed distribution
\begin{equation}
p(x_1,x_2,\ldots,x_N)\,=\,p_{x_1} p_{x_2} \cdots p_{x_N}\, =\,
p_1^{n_{\scriptscriptstyle 1}} p_2^{n_{\scriptscriptstyle 2}}\cdots
p_k^{n_{\scriptscriptstyle k}}\;.
\label{eq-iid}
\end{equation}
The numbers $p_j$ describe the objective, ``true'' probability that
the result of a single experiment will be $j$ ($j=1,\ldots,k$).  The
variable $n_j$, on the other hand, describes the number of times
outcome $j$ is listed in the vector $(x_1,x_2,\ldots,x_N)$. But this
description---for the objectivist---only describes the situation from
a kind of ``God's eye'' point of view.  To the experimentalist, the
``true'' probabilities $p_1,\ldots,p_k$ will very often be {\it
unknown\/} at the outset.  Thus, his burden is to estimate the
unknown probabilities by a statistical analysis of the experiment's
outcomes.

In the Bayesian approach, however, it does not make sense to talk
about estimating a true probability. Instead, a Bayesian assigns a
prior probability distribution $p(x_1,x_2,\ldots,x_N)$ on the
multi-trial hypothesis space and uses Bayes' theorem to update the
distribution in the light of his measurement results. The content of
de Finetti's theorem is this. Assuming {\it only\/} that
\be
p(x_{\pi(1)},x_{\pi(2)},\ldots,x_{\pi(N)}) = p(x_1,x_2,\ldots,x_N)
\ee
for any permutation $\pi$ of the set $\{1,\ldots,N\}$, and that for
any integer $M>0$, there is a distribution
$p_{N+M}(x_1,x_2,\ldots,x_{N+M})$ with the same permutation property
such that
\be
p(x_1,x_2,\ldots,x_N)\; =
\sum_{x_{N+1},\ldots,x_{N+M}}p_{N+M}(x_1,\ldots,x_N,x_{N+1},\ldots,x_{N+M})
\;,
\label{eq-marginal}
\ee
then $p(x_1,x_2,\ldots,x_N)$ can be written uniquely in the form
\bea
p(x_1,x_2,\ldots,x_N)&=& \int_{{\cal S}_k} P(\vec{p}\,)\,p_{x_1}
p_{x_2} \cdots p_{x_N}\,d\vec{p}
\label{Hestia}
\nonumber
\\
&=&
\int_{{\cal S}_k} P(\vec{p}\,)\, p_1^{n_{\scriptscriptstyle 1}}
p_2^{n_{\scriptscriptstyle 2}}\cdots p_k^{n_{\scriptscriptstyle k}}
\, d\vec{p}\;,
\label{eq-repr}
\eea
where $\vec{p}=(p_1,p_2,\ldots,p_k)$, and the integral is taken over
the simplex of such distributions
\be
{\cal S}_k=\left\{\vec{p}\mbox{ : }\; p_j\ge0\mbox{ for all } j\mbox{
and } \sum_{j=1}^k p_j=1\right\}.
\ee
Furthermore, the function $P(\vec{p}\,)\ge0$ is required to be a
probability density function on the simplex:
\begin{equation}
\int_{{\cal S}_k} P(\vec{p}\,)\,d\vec{p}=1\;,
\end{equation}
With this representation theorem, the unsatisfactory concept of an
unknown probability vanishes from the description in favor of the
fundamental idea of assigning an exchangeable probability
distribution to multiple trials.

This cue in hand, it is easy to see how to reword the description of
quantum-state tomography to meet our goals.  What is relevant is
simply a judgment on the part of the experimenter---notice the
essential subjective character of this ``judgment''---that there is
no distinction between the systems the device is preparing.  In
operational terms, this is the judgment that {\it all the systems
are and will be the same as far as observational predictions are
concerned}.  At first glance this statement might seem to be
contentless, but the important point is this: To make this
statement, one need never use the notion of an unknown state---a
completely operational description is good enough. Putting it into
technical terms, the statement is that if the experimenter judges a
collection of $N$ of the device's outputs to have an overall quantum
state $\rho^{(N)}$, he will also judge any permutation of those
outputs to have the same quantum state $\rho^{(N)}$. Moreover, he
will do this no matter how large the number $N$ is. This,
complemented only by the consistency condition that for any $N$ the
state $\rho^{(N)}$ be derivable from $\rho^{(N+1)}$, makes for the
complete story.

The words ``quantum state'' appear in this formulation, just as in
the original formulation of tomography, but there is no longer any
mention  of {\it unknown\/} quantum states.  The state $\rho^{(N)}$
is known by the experimenter (if no one else), for it represents his
state of knowledge.  More importantly, the experimenter is in a
position to make an unambiguous statement about the structure of the
whole sequence of states $\rho^{(N)}$: Each of the states
$\rho^{(N)}$ has a kind of permutation invariance over its factors.
The content of the {\it quantum de Finetti representation
theorem}~\cite{Hudson76,Caves01} is that a sequence of states
$\rho^{(N)}$ can have these properties, which are said to make it an
{\it exchangeable\/} sequence, if and only if each term in it can
also be written in the form
\begin{equation}
\rho^{(N)}=\int_{{\cal D}_d} P(\rho)\, \rho^{\otimes N}\, d\rho\;,
\label{Jeremy}
\end{equation}
where
\begin{equation}
\rho^{\otimes N}=
\underbrace{\rho\otimes\rho\otimes\cdots\otimes\rho}_{
\matrix{\mbox{$N$-fold tensor}\cr\mbox{product}}}
\end{equation}
Here $P(\rho)\ge0$ is a fixed probability distribution over the
density operator space ${\cal D}_d$, and
\begin{equation}
\int_{{\cal D}_d} P(\rho)\,d\rho=1\;,
\end{equation}
where $d\rho$ is a suitable measure.

The interpretive import of this theorem is paramount. For it alone
gives a mandate to the term unknown state in the usual description of
tomography.  It says that the experimenter can act {\it as if\/} his
state of knowledge $\rho^{(N)}$ comes about because he knows there
is a ``man in the box,'' hidden from view, repeatedly preparing the
same state $\rho$.  He does not know which such state, and the best
he can say about the unknown state is captured in the probability
distribution $P(\rho)$.

The quantum de Finetti theorem furthermore makes a connection to the
overarching theme of Bayesianism stressed above.  It guarantees for
two independent observers---as long as they have a rather minimal
agreement in their initial beliefs---that the outcomes of a
sufficiently informative set of measurements will force a
convergence in their state assignments for the remaining
systems~\cite{Schack00}.  This ``minimal'' agreement is
characterized by a judgment on the part of both parties that the
sequence of systems is exchangeable, as described above, and a
promise that the observers are not absolutely inflexible in their
opinions.  Quantitatively, the latter means that though $P(\rho)$ may
be arbitrarily close to zero, it can never vanish.

This coming to agreement works because an exchangeable density
operator sequence can be updated to reflect information gathered
from measurements by a another quantum version of Bayes's rule for
updating probabilities~\cite{Schack00}. Specifically, if
measurements on $K$ systems yield results $D_K$, then the state of
additional systems is constructed as in Eq.~(\ref{Jeremy}), but
using an updated probability on density operators given by
\begin{equation}
P(\rho|D_K)={P(D_K|\rho)P(\rho)\over P(D_K)}\;.
\label{QBayes}
\end{equation}
Here $P(D_K|\rho)$ is the probability to obtain the measurement
results $D_K$, given the state $\rho^{\otimes K}$ for the $K$
measured systems, and
\be
P(D_K)=\int_{{\cal D}_d} P(D_K|\rho)\,P(\rho)\,d\rho
\ee
is the unconditional probability for the measurement results. For a
sufficiently informative set of measurements, as $K$ becomes large,
the updated probability $P(\rho|D_K)$ becomes highly peaked on a
particular state $\rho_{D_K}$ dictated by the measurement results,
regardless of the prior probability $P(\rho)$, as long as $P(\rho)$
is nonzero in a neighborhood of $\rho_{D_K}$. Suppose the two
observers have different initial beliefs, encapsulated in different
priors $P_i(\rho)$, $i=1,2$.  The measurement results force them to a
common state of knowledge in which any number $N$ of additional
systems are assigned the product state $\rho_{D_K}^{\otimes N}$,
i.e.,
\begin{equation}
\int P_i(\rho|D_K)\,\rho^{\otimes N}\,d\rho
\quad{\longrightarrow}\quad \rho_{D_K}^{\otimes N}\;,
\label{HannibalLecter}
\end{equation}
independent of $i$, for $K$ sufficiently large.

This shifts the perspective on the purpose of quantum-state
tomography:  It is not about uncovering some ``unknown state of
nature,'' but rather about the various observers' coming to
agreement over future probabilistic predictions. In this connection,
it is interesting to note that the quantum de Finetti theorem and
the conclusions just drawn from it work only within the framework of
complex vector-space quantum mechanics. For quantum mechanics based
on real Hilbert spaces, the connection between exchangeable density
operators and unknown quantum states does not hold.

A simple counterexample is the following.  Consider the $N$-system
state
\begin{equation}
\rho^{(N)}={1\over2}\rho_+^{\otimes N} + {1\over2}\rho_-^{\otimes N}
\;,
\label{eq-real}
\end{equation}
where
\begin{equation}
\rho_+={1\over2}(I+\sigma_2)\qquad\mbox{and}
\qquad\rho_-={1\over2}(I-\sigma_2)
\end{equation}
and $\sigma_1$, $\sigma_2$, and $\sigma_3$ are the Pauli matrices.
In complex-Hilbert-space quantum mechanics, Eq.~(\ref{eq-real}) is
clearly a valid density operator:  It corresponds to an equally
weighted mixture of $N$ spin-up particles and $N$ spin-down
particles in the $y$-direction.  The state $\rho^{(N)}$ is thus
exchangeable, and the decomposition in Eq.~(\ref{eq-real}) is unique
according to the quantum de Finetti theorem.

But now consider $\rho^{(N)}$ as an operator in real-Hilbert-space
quantum mechanics.  Despite its ostensible use of the imaginary
number $i$, it remains a valid quantum state.  This is because, upon
expanding the right-hand side of Eq.~(\ref{eq-real}), all the terms
with an odd number of $\sigma_2$'s cancel away.  Yet, even though it
is an exchangeable density operator, it cannot be written in de
Finetti form Eq.~(\ref{Jeremy}) using only real symmetric operators.
This follows because $i\sigma_2$ cannot be written as a linear
combination of $I$, $\sigma_1$, and $\sigma_3$, while a
real-Hilbert-space de Finetti expansion as in Eq.~(\ref{Jeremy}) can
{\it only\/} contain those three operators. Hence the de Finetti
theorem does not hold in real-Hilbert-space quantum mechanics.

In classical probability theory, exchangeability characterizes those
situations where the only data relevant for updating a probability
distribution are frequency data, i.e., the numbers $n_j$ in
Eq.~(\ref{eq-repr}). The quantum de Finetti representation shows
that the same is true in quantum mechanics: Frequency data (with
respect to a sufficiently robust measurement)\footnote{Technically,
this means any POVM $\{E_b\}$ whose elements span the space of
Hermitian operators. See Ref.~\cite{Caves01} for details.} are
sufficient for updating an exchangeable state to the point where
nothing more can be learned from sequential measurements. That is,
one obtains a convergence of the form Eq.~(\ref{HannibalLecter}), so
that ultimately any further measurements on the individual systems
will be statistically independent. That there is no quantum de
Finetti theorem in real Hilbert space means that there are
fundamental differences between real and complex Hilbert spaces with
respect to learning from measurement results.

Finally, in summary, let us hang on the point of learning for just a
little longer.  The quantum de Finetti theorem shows that the
essence of quantum-state tomography is not in revealing an ``element
of reality'' but in deriving that various agents (who agree some
minimal amount) can come to agreement in their ultimate quantum-state
assignments. This is not the same thing as the stronger statement
that ``reality does not exist.''  It is simply that one need not go
to the extreme of taking the ``unknown quantum state'' as being
objectively real to make sense of the experimental practice of
tomography.

One is left with the feeling---an almost salty
feeling\footnote{Working under the presumption that no interpretation
of quantum mechanics is worth its salt unless it raises as many
technical questions as it answers philosophical ones.}---that
perhaps this is the whole point to quantum mechanics. That is:
Perhaps the missing ingredient for narrowing the structure of
Bayesian probability down to the structure of quantum mechanics has
been in front of us all along.  It finds no better expression than in
the taking account of the limitations the physical world poses to our
ability to come to agreement.

\section{The Oyster and the Quantum}
\begin{flushright}
\parbox{4.0in}{\footnotesize
\bq
\indent The significance of this development is to give us insight into
the logical possibility of a new and wider pattern of thought.  This
takes into account the observer, including the apparatus used by him,
differently from the way it was done in classical physics \ldots\ In
the new pattern of thought we do not assume any longer the {\it
detached observer},occurring in the idealizations of this classical
type of theory, but an observer who by his indeterminable effects
creates a new situation, theoretically described as a new state of
the observed system. \ldots

Nevertheless, there remains still in the new kind of theory an {\it
objective reality}, inasmuch as these theories deny any possibility
for the observer to influence the results of a measurement, once the
experimental arrangement is chosen. Particular qualities of an
individual observer do not enter the conceptual framework of the
theory.
\\
\hspace*{\fill} --- {\it Wolfgang Pauli}, 1954
\eq
}
\end{flushright}

A grain of sand falls into the shell of an oyster and the result is
a pearl.  The oyster's sensitivity to the touch is the source of one
of our most beautiful gems. In the 75 years that have passed since
the founding of quantum mechanics, only the last 10 have turned to a
view and an attitude that may finally reveal the essence of the
theory.  The quantum world is sensitive to the touch, and that may
well be one of the best things about it.  Quantum information
science---with its three prongs of quantum information theory,
quantum cryptography, and quantum computing---leads the way in
telling us how to quantify that sentence.  Quantum algorithms can be
exponentially faster than classical algorithms.  Secret keys can be
encoded into physical systems in such a way as to reveal whether
information has been gathered about them.  The list of triumphs
keeps growing.

The key to so much of this has been simply in a change of attitude.
This can be seen by going back to almost any older textbook on
quantum mechanics: Nine times out of ten, the Heisenberg uncertainty
relation is presented in a way that conveys the feeling that we have
been short-changed by the physical world.

\pagebreak

\bq
\small
``Look at classical physics, how nice it is:  We can measure a
particle's position and momentum with as much accuracy as we would
like.  How limiting quantum theory is instead.  We are stuck with
$$
\Delta x \Delta p \ge \frac{1}{2}\hbar\;,
$$
and there is just nothing we can do about it. The task of physics is
just to sober up to this state of affairs and make the best of it.''
\eq
How this contrasts with the point of departure of quantum information
science!  There the task is not to ask what limits quantum mechanics
places upon us, but what novel, productive things we can do in the
quantum world that we could not have done otherwise. In what ways is
the quantum world fantastically better than the classical one?

If one is looking for something ``real'' in quantum theory, what more
direct tack could one take than to look to its technologies? People
may argue about the objective reality of the wave function ad
infinitum, but few would argue about the existence of quantum
cryptography as a solid prediction of the theory. Why not take that
or a similar effect as the grounding for what quantum mechanics is
trying to tell us about nature?

Let us try to give this imprecise set of thoughts some shape by
molding quantum cryptography into the vision built up in the previous
sections.  For quantum key distribution it is essential to be able
to prepare a physical system in one or another quantum state drawn
from some fixed {\it nonorthogonal\/} set
\cite{Bennett84,Bennett92}. These nonorthogonal states are used to
encode a potentially secret cryptographic key to be shared between
the sender and receiver. The information an eavesdropper seeks is
about which quantum state was actually prepared in each individual
transmission. What is novel here is that the encoding of the
proposed key into nonorthogonal states forces the
information-gathering process to induce a disturbance to the overall
{\it set\/} of states. That is, the presence of an active
eavesdropper transforms the initial pure states into a set of mixed
states or, at the very least, into a set of pure states with larger
overlaps than before. This action ultimately boils down to a loss of
predictability for the sender over the outcomes of the receiver's
measurements and, so, is directly detectable by the receiver (who
reveals some of those outcomes for the sender's inspection). More
importantly, there is a direct connection between the statistical
information gained by an eavesdropper and the consequent disturbance
she must induce to the quantum states in the process.  As the
information gathered goes up, the necessary disturbance also goes up
in a precisely formalizable way \cite{Fuchs96}.

Note the two ingredients that appear in this scenario. First, the
information gathering or measurement is grounded with respect to one
observer (in this case, the eavesdropper), while the disturbance is
grounded with respect to another (here, the sender). In particular,
the disturbance is a disturbance to the sender's previous
information---this is measured by her diminished ability to predict
the outcomes of certain measurements the legitimate receiver might
perform. No hint of any variable intrinsic to the system is made use
of in this formulation of the idea of ``measurement causing
disturbance.''

The second ingredient is that one must consider at least two possible
non\-orthogonal preparations in order for the formulation to have
any meaning. This is because the information gathering is not about
some classically-defined observable---i.e., about some unknown
hidden variable or reality intrinsic to the system---but is instead
about which of the unknown states the sender actually prepared.  The
lesson is this: Forget about the unknown preparation, and the random
outcome of the quantum measurement is information about nothing.  It
is simply ``quantum noise'' with no connection to any preexisting
variable.

How crucial is this second ingredient---that is, that there be at
least two nonorthogonal states within the set under consideration?
We can address its necessity by making a shift in the account above:
One might say that the eavesdropper's goal is not so much to uncover
the identity of the unknown quantum state, but to sharpen her
predictability over the receiver's measurement outcomes.  In fact,
she would like to do this at the same time as disturbing the
sender's predictions as little as possible.  Changing the language
still further to the terminology of Section 4, the eavesdropper's
actions serve to sharpen her information about the potential
consequences of the receiver's further interventions on the system.
(Again, she would like to do this while minimally diminishing the
sender's previous information about those same consequences.) In the
cryptographic context, a byproduct of this effort is that the
eavesdropper ultimately comes to a more sound prediction of the
secret key. From the present point of view, however, the importance
of this change of language is that it leads to an almost Bayesian
perspective on the information--disturbance problem.

As previously emphasized, within Bayesian probability the most
significant theme is to identify the conditions under which a set of
decision-making agents can come to a common probability assignment
for some random variable in spite of the fact that their initial
probabilities differ \cite{Bernardo94}.  One might similarly view
the process of quantum eavesdropping.  The sender and the
eavesdropper start off initially with differing quantum state
assignments for a single physical system.  In this case it so
happens that the sender can make sharper predictions than the
eavesdropper about the outcomes of the receiver's measurements.  The
eavesdropper, not satisfied with this situation, performs a
measurement on the system in an attempt to sharpen those
predictions.  In particular, there is an attempt to come into
something of an agreement with the sender but without revealing the
outcomes of her measurements or, indeed, her very presence.

It is at this point that a distinct {\it property\/} of the quantum
world makes itself known.  The eavesdropper's attempt to
surreptitiously come into alignment with the sender's predictability
is always shunted away from its goal.  This shunting of various
observer's predictability is the subtle manner in which the quantum
world is sensitive to our experimental interventions.

And maybe this is our crucial hint!  The wedge that drives a
distinction between Bayesian probability theory in general and
quantum mechanics in particular is perhaps nothing more than this
``Zing!''~of a quantum system that is manifested when an agent
interacts with it. It is this wild sensitivity to the touch that
keeps our knowledge and beliefs from ever coming into too great of
an alignment. The most our knowledge about the potential
consequences of our interventions on a system can come into
alignment is captured by the mathematical structure of a pure
quantum state $|\psi\rangle$.  Take all possible
information-disturbance curves for a quantum system, tie them into a
bundle, and {\it that\/} is the long-awaited property, the input we
have been looking for from nature.

Or, at least, that is the speculation.  Look at that bundle long and
hard and we might just find that it stays together without the help
of our tie.

\section{Acknowledgments}

I thank Carl Caves, Greg Comer, and R\"udiger Schack for inspiration
and years of conversation.  I further thank Steven van Enk, David
Mermin, and Asher Peres for suggestions on the manuscript, and Sam
Braunstein for trying to turn my e's upside down.  This work was
performed in part during the program on ``Quantum Measurement and
Information'' at the Erwin Schr\"odinger International Institute for
Mathematical Physics in Vienna.

\end{document}